\documentclass[useAMS,usenatbib]{mn2e}
\usepackage{epsfig}
\usepackage{graphicx}
\usepackage{times}
\usepackage{color}

\def\sun{\hbox{$\odot$}}
\def\cm3{cm$^{-3}$}
\def\kms{km~s$^{-1}$}

\def\msun{M$_{\odot}$}

\def\beq{\begin{equation}}
\def\eeq{\end{equation}}

\def\lesssim{\mathrel{\hbox{\rlap{\hbox{\lower4pt\hbox{$\sim$}}}\hbox{$<$}}}}
\def\gtrsim{\mathrel{\hbox{\rlap{\hbox{\lower4pt\hbox{$\sim$}}}\hbox{$>$}}}}

\def\lesssim{\mathrel{\hbox{\rlap{\hbox{\lower4pt\hbox{$\sim$}}}\hbox{$<$}}}}
\def\gtrsim{\mathrel{\hbox{\rlap{\hbox{\lower4pt\hbox{$\sim$}}}\hbox{$>$}}}}

\def\one{\,{\,\sc i}}
\def\two{\,{\,\sc ii}}
\def\three{\,{\,\sc iii}}

\newcommand{\gcc}{\ensuremath{\rm{g\,cm}^{-3}}}

\def\cmfgen{{\sc cmfgen}}

\def\nifs{\iso{56}Ni}

\newcommand{\iso}[2]{\ensuremath{^{#1}\rm{#2}}}

\def\aj{AJ}
\def\pasp{PASP}
\def\apj{ApJ}

\def\apjl{ApJL}
\def\aap{A\&A}

\def\mnras{MNRAS}
\def\nat{Nature}

\title[Constraints on SN Ia explosions]{Constraints on the explosion mechanism and
progenitors of type Ia supernovae}

\author[Luc Dessart, St\'ephane Blondin, D.J. Hillier, and Alexei Khokhlov]
{Luc Dessart,$^{1,2}$ St\'ephane Blondin,$^{1}$ D. John Hillier,$^{3}$ and Alexei Khokhlov$^{4}$ \\ \\
$^{1}$ Aix Marseille Universit\'e, CNRS, LAM (Laboratoire d'Astrophysique
de Marseille) UMR 7326, 13388, Marseille, France.\\
$^{2}$ Laboratoire Lagrange, UMR7293, Universit\'e Nice Sophia-Antipolis, CNRS,
Observatoire de la C\^{o}te d'Azur, 06300 Nice, France. \\
$^{3}$: Department of Physics and Astronomy \& Pittsburgh Particle physics, Astrophysics,
and Cosmology Center (PITT PACC), University of Pittsburgh,   \\
3941 O'Hara Street, Pittsburgh, PA 15260, USA. \\
$^4$ Department of Astronomy \& Astrophysics, the Enrico Fermi Institute, and the Computational Institute,
The University of Chicago, \\
Chicago, IL 60637, USA}

\begin{document}

\date{Accepted 2014 March 24.  Received 2014 March 21; in original form 2013 October 29}

\pagerange{\pageref{firstpage}--\pageref{lastpage}} \pubyear{2014}

\maketitle

\label{firstpage}

\begin{abstract}
Observations of SN 2011fe at early times reveal an evolution analogous to a fireball model of constant color.
In contrast, our unmixed delayed detonations of Chandrasekhar-mass white dwarfs (DDC series)
exhibit a faster brightening concomitant with a shift in color to the blue.
In this paper, we study the origin of these discrepancies.
We find that strong chemical mixing largely resolves the photometric mismatch at early times, 
but it leads to an enhanced line broadening that contrasts, for example, 
with the markedly narrow Si\two\,6355\,\AA\ line  of SN\,2011fe.
We also explore an alternative configuration with pulsational-delayed  detonations (PDDEL model series).
Because of the pulsation, PDDEL models retain more unburnt carbon,
have little mass at high velocity, and have a much hotter outer ejecta after the explosion.
The pulsation does not influence the inner ejecta, so PDDEL and DDC models exhibit similar
radiative properties beyond maximum.
However, at early times, PDDEL models show bluer optical colors and a higher luminosity, even for weak mixing.
Their early-time radiation is derived primarily from the initial shock-deposited energy in the outer ejecta
rather than radioactive decay heating. Furthermore, PDDEL models show short-lived C\two\ lines,
reminiscent of SN\,2013dy. They typically exhibit lines that are weaker, narrower, and of near-constant width, 
reminiscent of SN\,2011fe.
In addition to multi-dimensional effects, varying configurations for such ``pulsations" offer a source
of spectral diversity amongst SNe Ia.
PDDEL and DDC models also provide one explanation for low- and high-velocity gradient SNe Ia.
\end{abstract}


\section{Introduction}

Supernovae (SNe) Ia likely result from the explosion of carbon-oxygen degenerate stars
in binary systems \citep{hoyle_fowler_60}.
At present, the explosion mechanism known as the delayed-detonation model
seems to offer the best agreement with SN Ia observations, provided one varies
the deflagration-to-detonation transition density to yield a range of \nifs\ mass.
Successes of this model include the reproduction of the observed range of peak
luminosities; the correct stratification of chemical elements in
the ejecta to match the spectral line widths of C, O, intermediate-mass elements (IMEs) and
iron-group elements (IGEs); the proper correlation between maximum brightness and width of
the light curve \citep{khokhlov_etal_93,hoeflich_khokhlov_96,hoeflich_etal_96,hoeflich_etal_10};
the diversity in maximum-light spectra from sub-luminous to ``standard" luminosity events
\citep{blondin_etal_13}.  Modeling of X-ray spectra from SN Ia remnants provides independent
support for this delayed-detonation mechanism
\citep{badenes_etal_06,badenes_etal_08,patnaude_etal_12,ciotellis_etal_13}.

In \citet[hereafter D13]{dessart_etal_13xx}, we presented an investigation, probably not exhaustive,
of critical ingredients for the radiative-transfer modeling of SN Ia ejecta until the onset of the nebular
phase. Our work revealed the need
to treat a variety of non-LTE processes that are crucial to obtain reliable
temperature and ionization structures. With such ingredients included, it appears that one can
capture the fundamental spectral characteristics of SNe Ia from $-$10 to +40\,d after $B$-band maximum,
i.e., at least until the onset of the nebular phase. How well the delayed-detonation scenario reproduces
the observed evolution of SNe Ia during their first week after explosion has been less studied. Interest in this question
is warranted today by the novel observational constraints set by the nearby SN\,2011fe
\citep{nugent_etal_11,richmond_etal_12,parrent_etal_12,pereira_etal_13}.

The color and brightness of a SN Ia at early times may be affected in various ways,
but the consensus is that it is controlled by \nifs\ decay heating
\citep{colgate_mckee_69,arnett_82,piro_12}.
Multi-D fluid instabilities inherent to combustion may produce a complex 3-D
chemical distribution in the ejecta, and in particular mixing of \nifs-rich
material that could influence the outer ejecta \citep{gamezo_etal_05}, its heat content
and therefore the SN Ia radiation at early times.
In numerical simulations, ejecta asphericity seem further enhanced, perhaps unrealistically,
by a non-spherical initiation of the explosion itself
\citep{livne_etal_05,roepke_hillebrandt_05}.\footnote{Diversity in progenitor properties may also be
relevant, e.g., peculiarities from the simmering phase, asphericity due to rotation,
presence of an accretion disk etc.}
Such ejecta asphericities cause the SN radiation to be angle dependent, producing
one possible explanation to SN spectral diversity \citep{kasen_etal_09,maeda_etal_10}.
Observationally, SNe Ia exhibit diverse trajectories for the P-Cygni
profile minima in velocity space, implying distinct recession velocities
of SN Ia ejecta photospheres \citep{benetti_etal_05,blondin_etal_12}.

Early time observations are probably best to reveal the diversity of progenitor and explosion
properties since early-time spectra and photometry should show signatures related to the companion star \citep{kasen_10},
or the structure of the outer layers of the white dwarf as a result of a merger and/or complex accretion processes.
However, there  has been only a limited discussion in the literature concerning the specific behavior of delayed-detonation
models at early times, and then they are usually limited to semi-analytical solutions
for the luminosity (e.g., \citealt{piro_12}).
Since SNe Ia may exhibit a ``dark phase" \citep{piro_Nakar_13}, there is also some tension in the
explosion time inferred from observations and models \citep{mazzali_etal_13}, suggesting that
such a fundamental quantity still retains some uncertainty even for the best observed Type Ia SN, SN 2011fe.
For that SN, \citet{roepke_etal_12} primarily focus  on the maximum-light spectrum to discriminate
between explosion models.
Previous time-dependent simulations of SNe Ia, generally performed
using the Monte Carlo technique, typically adopt a modest maximum grid velocity of $\sim$\,25000\,\kms\
(e.g., \citealt{kasen_etal_09}).
This value is somewhat small for a radiative-transfer study at  early post-explosion time, when the spectrum-formation region
is located in the outermost ejecta layers. Poor photon statistics also impact the accuracy of predicted observables.
This can prevent an assessment of the properties of weaker lines, such as C\two\ line emission/absorption
(see, e.g., \citealt{sim_etal_10}) which can carry critical information on the explosion mechanism.

It is commonly assumed that the light-curve of Type Ia SNe is almost entirely controlled by \nifs\
through heating of the ejecta --- any observational signature of the initial temperature arising from
the SN explosion being lost (before the ejecta is 1 day old) due to the large expansion rate.  When
there are discrepancies of the light-curve with observations, mixing or ejecta asymmetry is invoked.
However it is necessary to consider
whether the outer temperature of SN Ia ejecta could be larger than that generally obtained in
standard explosions of Chandrasekhar-mass white dwarfs  in hydrostatic equilibrium.
In Nature, a larger progenitor radius (as in the extreme case of red-supergiant stars)
can cause a SN to remain more luminous and relatively bluer for longer.
In the case of a white dwarf, very large variations would be needed to sufficiently swell the star
and quench cooling from expansion. Such modulations are therefore unlikely to be large
enough to produce the desired effect. Instead, an attractive configuration is an
explosion from a pre-expanded white dwarf or from a white dwarf with a buffer of mass around it, or similar
configurations that may arise from a pulsation or from a binary merger.
As we show in this paper, such configurations will alter in other ways the outer ejecta
and offer an interesting means to generate spectral diversity from SN Ia models,
even for a quasi spherical ejecta.

In the next section, we present the observational data we use to confront with our models.
We then investigate the behavior of our delayed-detonation models at early times, and in particular
discuss the ingredients that affect their properties (Section~\ref{sect_pb}).
The potential merits of pulsational-delayed detonation models are explored in  Section~\ref{sect_setup}.
In this section we also compare the pulsational-delayed detonation models with observations of the type Ia SN\,2011fe.
We present the radiative properties that differ between pulsational-delayed detonation models
and ``standard" delayed-detonation models in Section~\ref{sect_res}.
Finally, we present our conclusions in Section~\ref{sect_conc}.

\begin{figure}
\epsfig{file=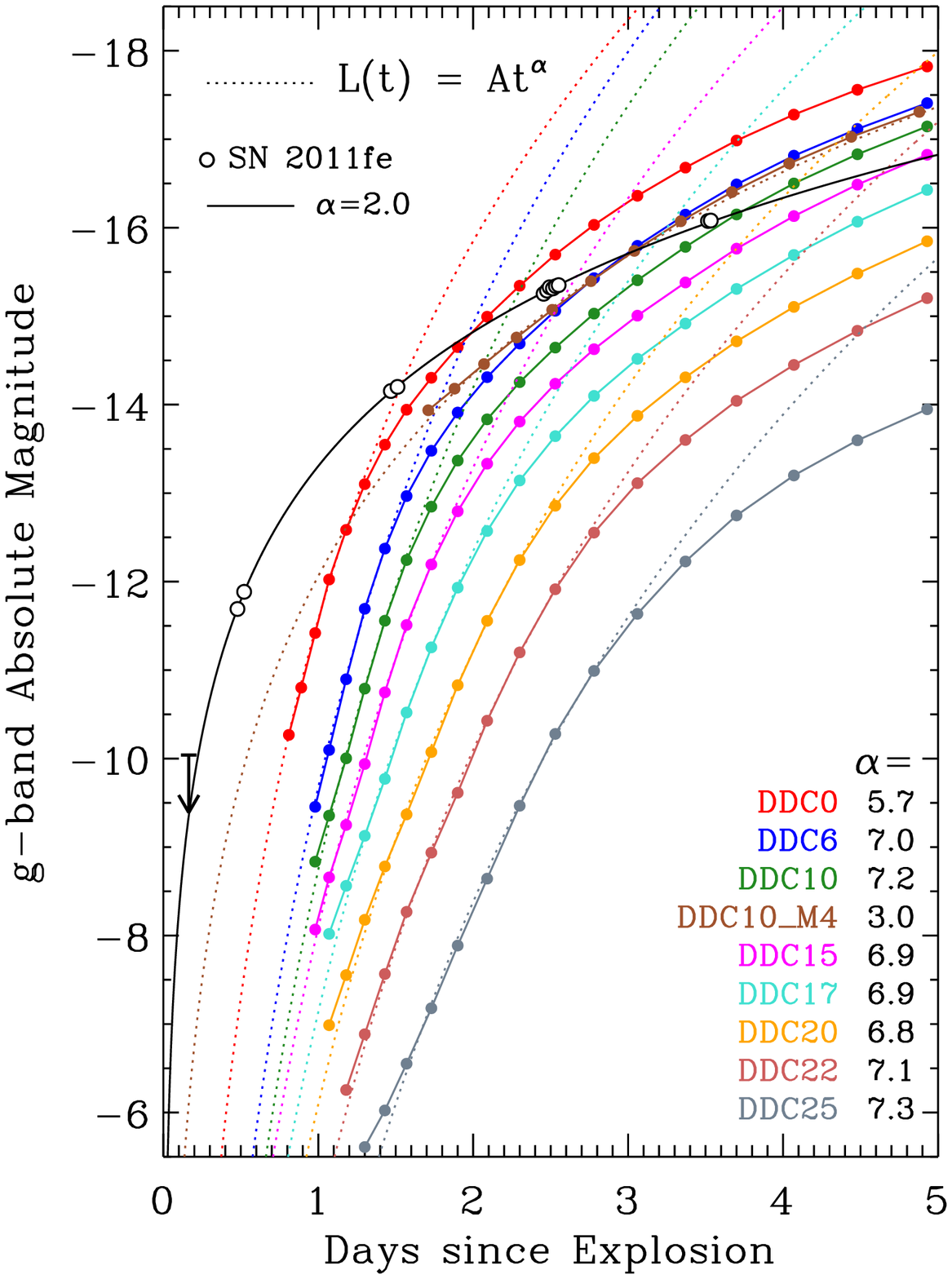,width=8.75cm}
\epsfig{file=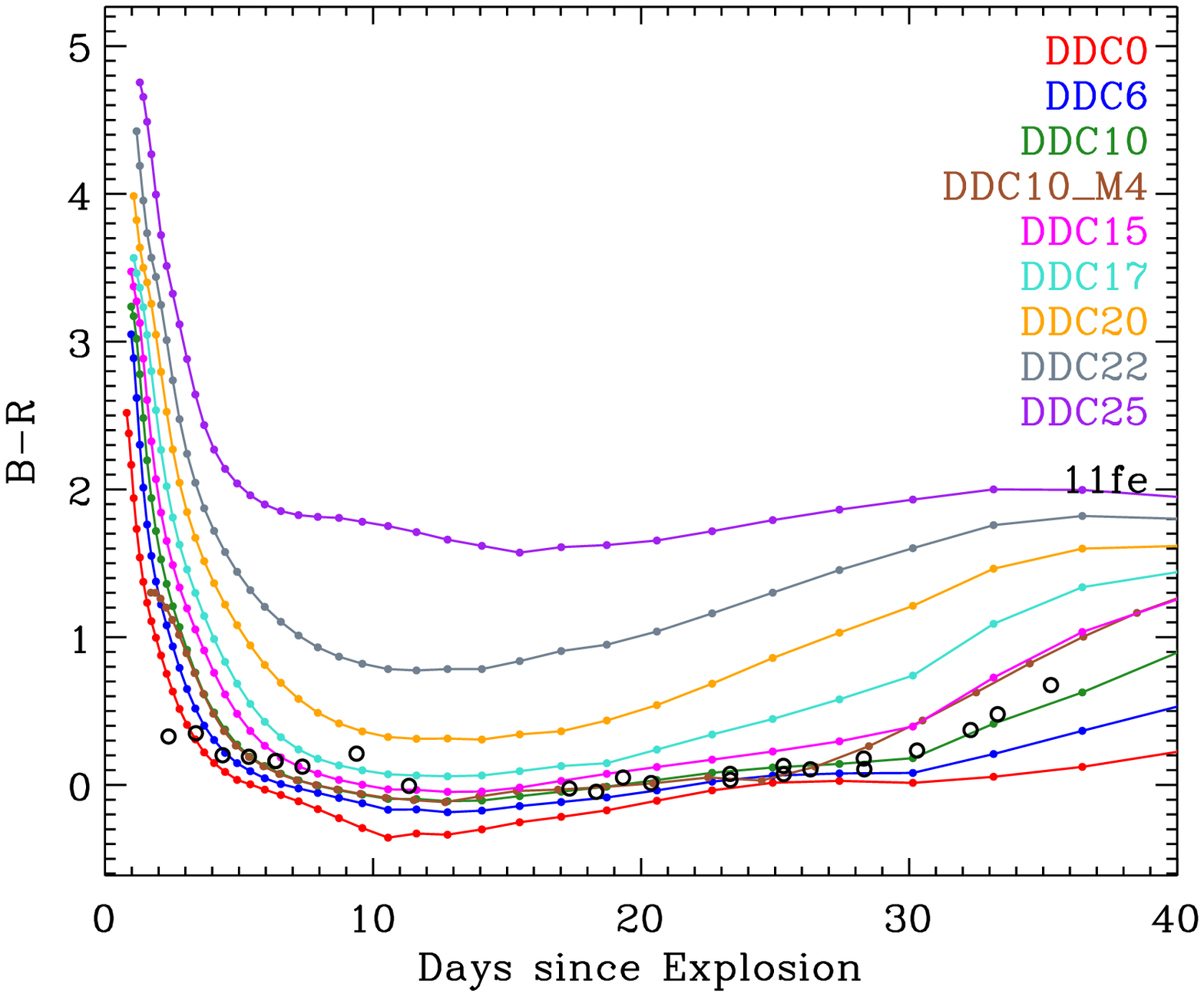,width=8.75cm}
\caption{{\bf Top:} Absolute $g$-band light curve for the DDC delayed-detonation models (connected colored lines
with filled symbols, which correspond to the steps in each time sequence) presented in \citet{blondin_etal_13},
together with the observed $g$-band light curve of SN\,2011fe (\citealt{nugent_etal_11}; we adopt
their explosion time). The upper limit placed on the magnitude prior to detection is from \citet{bloom_etal_12}.
While a power-law with exponent $\alpha=$2 matches the quasi constant color evolution of SN\,2011fe,
DDC models have steeper slopes at early times ($\alpha$ on the order of 6--7) and are systematically too faint
prior to 2--3\,d, irrespective of \nifs\ mass. The discrepancy remains if the adopted explosion time is shifted by
a day or so.
{\bf Bottom:} $B-R$ color evolution of the models shown at top, together with the corresponding color
of SN\,2011fe (opened symbols; \citealt{richmond_etal_12}).
Besides being sub-luminous, our DDC models are significantly redder prior to 2--3\,d.
\label{fig_slope}
}
\end{figure}

\section{Sources of observational data}
\label{sect_obs}

To test the broad compatibility of our \cmfgen\ simulations
we primarily compare them to the multi-epoch observations of SN\,2011fe. We take the photometric data from
\citet{richmond_etal_12} and the observed spectra from \citet{pereira_etal_13}. As in \citet{nugent_etal_11},
we assume no reddening and adopt a distance modulus of 29.04\,mag \citep{shappee_stanek_11}.
We use the host-galaxy redshift  of 0.00089 quoted by
NED ($cz=267\pm20$\,\kms\ from the Updated Zwicky Catalogue; \citealt{falco_etal_99}).
We adopt the inferred explosion time of \citet{nugent_etal_11}, namely MJD\,55796.687.

For some comparisons we also use observational data for SN\,2002bo from \citet{benetti_etal_04},
and for SN\,2005cf from \citet{garavini_etal_07} and \citet{bufano_etal_09}.

\begin{table}
\caption{Summary of model nomenclature and properties for the \cmfgen\ simulations
discussed in Section~\ref{sect_ddc_mix}-\ref{sect_nuc}.
All simulations are based on the  delayed-detonation model named DDC10
(see D13 and \citealt{blondin_etal_13} for details).
\label{tab_modset_DDC}}
\begin{center}
\begin{tabular}{lc@{\hspace{2.mm}}c@{\hspace{2.mm}}
c@{\hspace{2mm}}c@{\hspace{2.mm}}}
\hline\hline
Model &  M(\nifs) & Start time & Decay chains   &    $v_{\rm mix}$      \\
      &   [\msun] &  [d]      &                &    [\kms]                 \\
\hline
DDC10\_M0     & 0.623  &  1   & \iso{56}Ni only   & 0      \\
DDC10\_M1      & 0.623 &  1   & \iso{56}Ni only   & 250   \\
DDC10\_M2     & 0.623 &  1   & \iso{56}Ni only    & 500   \\
DDC10\_M3     & 0.623 &  1   & \iso{56}Ni only    & 1000  \\
DDC10\_M4     & 0.623  &  1   & \iso{56}Ni only   & 1500    \\
\hline
DDC10\_T1D0   & 0.623  &  1     & \iso{56}Ni only   & 500     \\
DDC10\_T0D2   & 0.623  & 0.5   & 1-step \& 2-step  & 500     \\
\hline
\end{tabular}
\end{center}
\end{table}

\section{Early-time behavior: Dependencies}
\label{sect_pb}

   The procedure used to compute SN light curves and spectra with
\cmfgen\ has been described numerous times
\citep{dessart_etal_12,HD12,dessart_etal_13b,dessart_etal_13}
and will not be repeated here. In this work on
SNe Ia, we use the same approach as discussed in \citet{blondin_etal_13} and D13.
Model atoms used are A1 up to the peak and A4 beyond (see appendix of
D13).

 In \cmfgen, we adopt a time step equal to 10\% of
the current time, and generally start the simulations at about 1\,d. The problem is that the
initial relaxation from the initial model takes a few time steps, and thus
compromises the reliability of our models at 1\,d (this relaxation is unavoidable because
we start from a pure hydro model etc.). This is problematic if we
want to compare our models to early-time observations of SN\,2011fe.
We therefore start most of the \cmfgen\ simulations in this work at 0.5\,d after explosion.
When showing results, we omit the first computed time steps to skip the
``relaxation" stage.

\begin{figure}
\epsfig{file=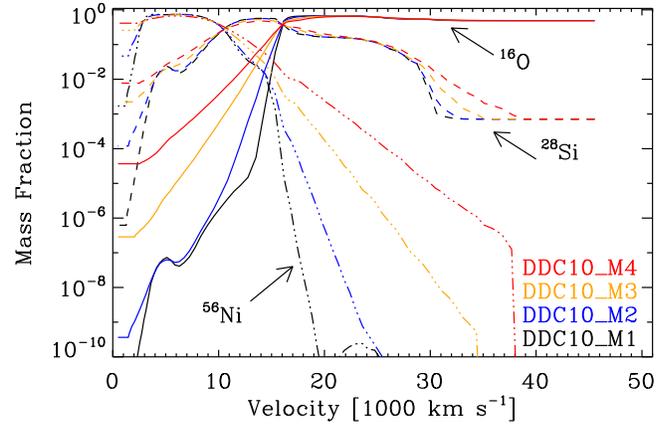,width=8.5cm}
\caption{Chemical stratification of DDC10 ejecta models in which some radial mixing
has been applied. We only show the representative species \iso{16}O (solid),
\iso{28}Si (dashed), and \iso{56}Ni (dash-dotted; the post-explosion time is 1.2\,d).
The ``boxcar" velocity we use for mixing is 250, 500, 1000, and 1500\,\kms\ from models M1 to M4. 
For the weakest radial mixing (model M1), the stratification is essentially
the same as the distribution of the original, un-mixed, ejecta model.
\label{fig_init_mix}
}
\end{figure}

\begin{figure*}
\centering
\begin{flushleft}
\begin{minipage}{0.5\linewidth}
\epsfig{file=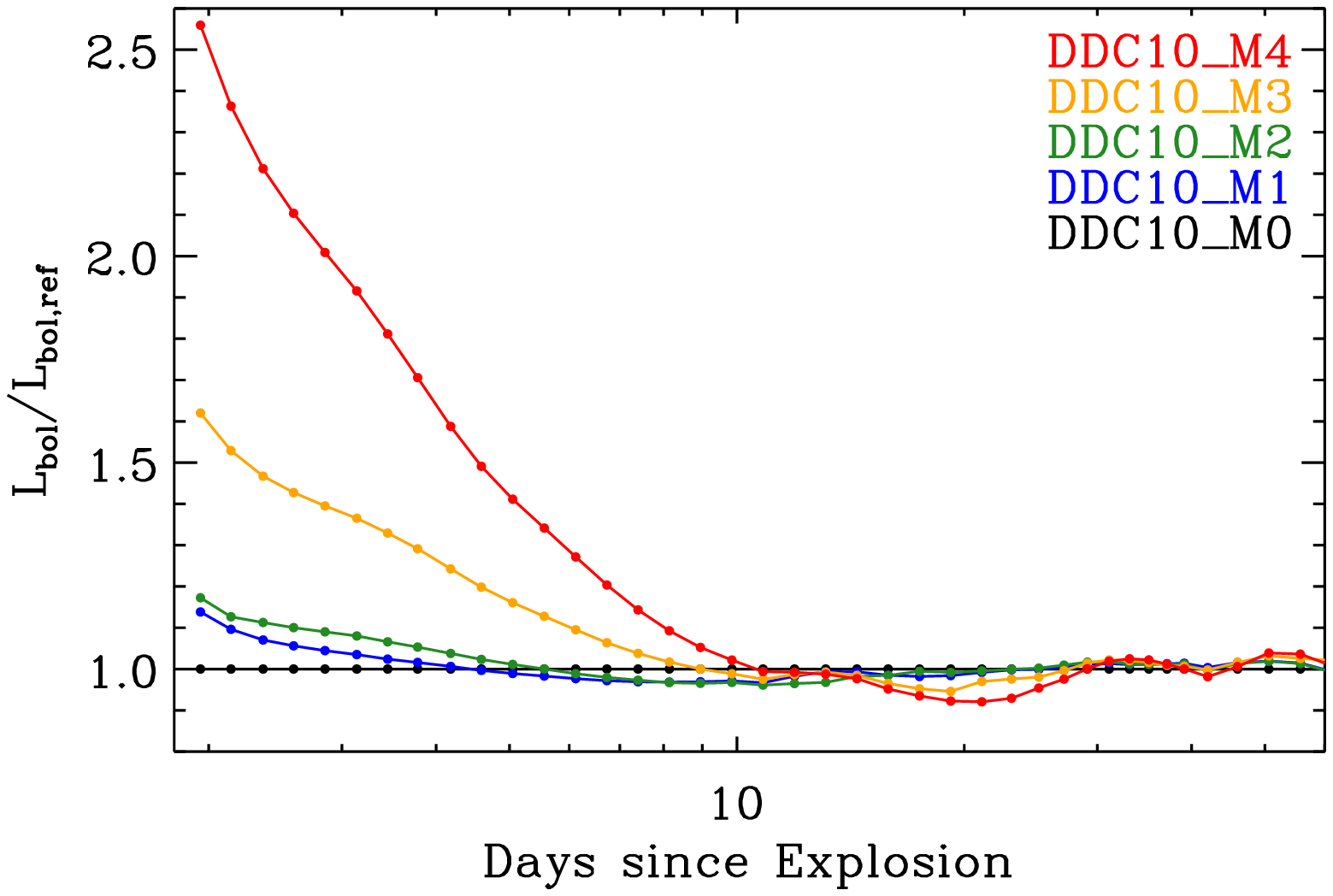,width=8.5cm}
\epsfig{file=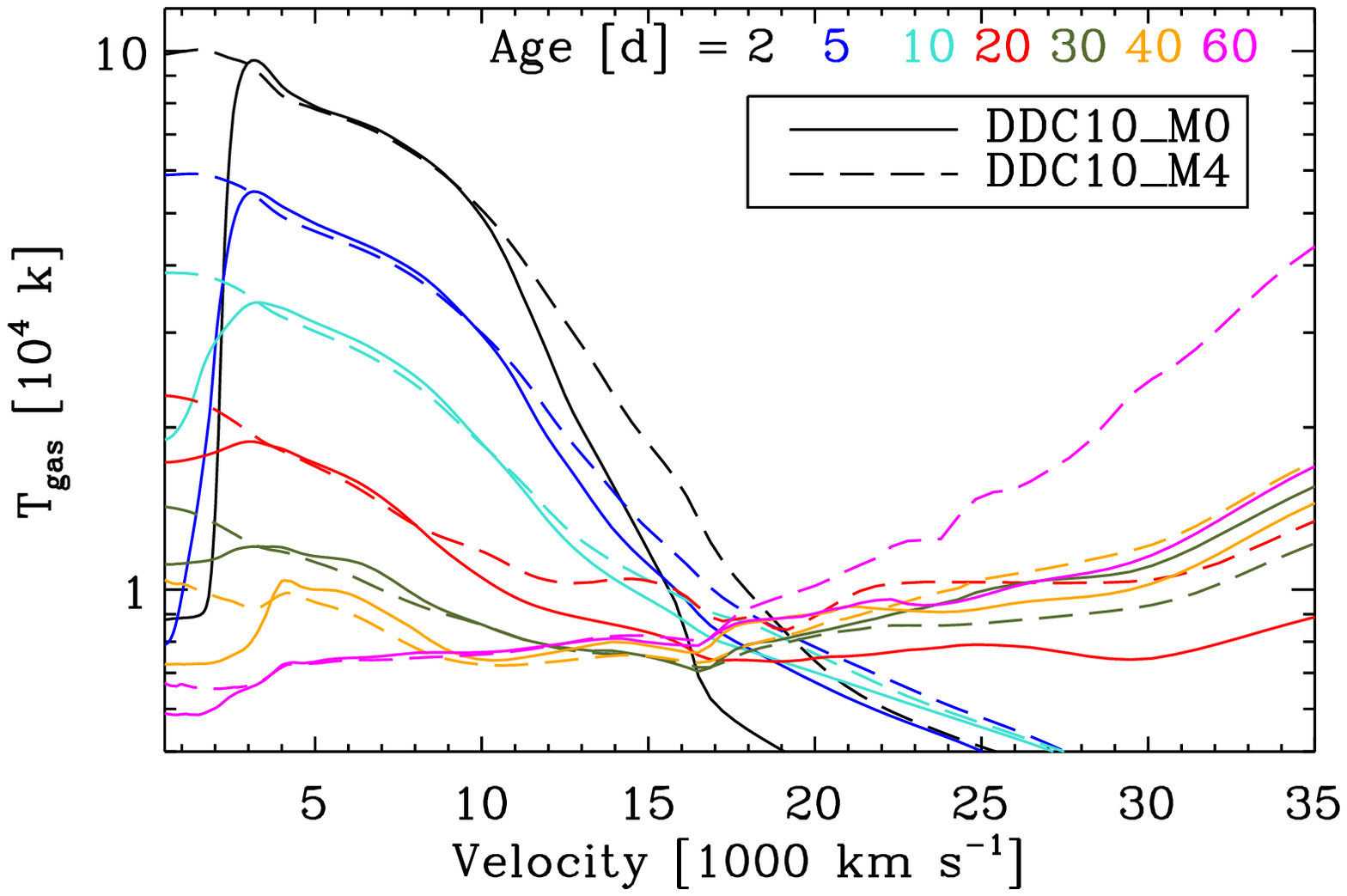,width=8.5cm}
\end{minipage}
\end{flushleft}
\vspace{-11.8cm}
\begin{flushright}
\begin{minipage}{0.5\linewidth}
\epsfig{file=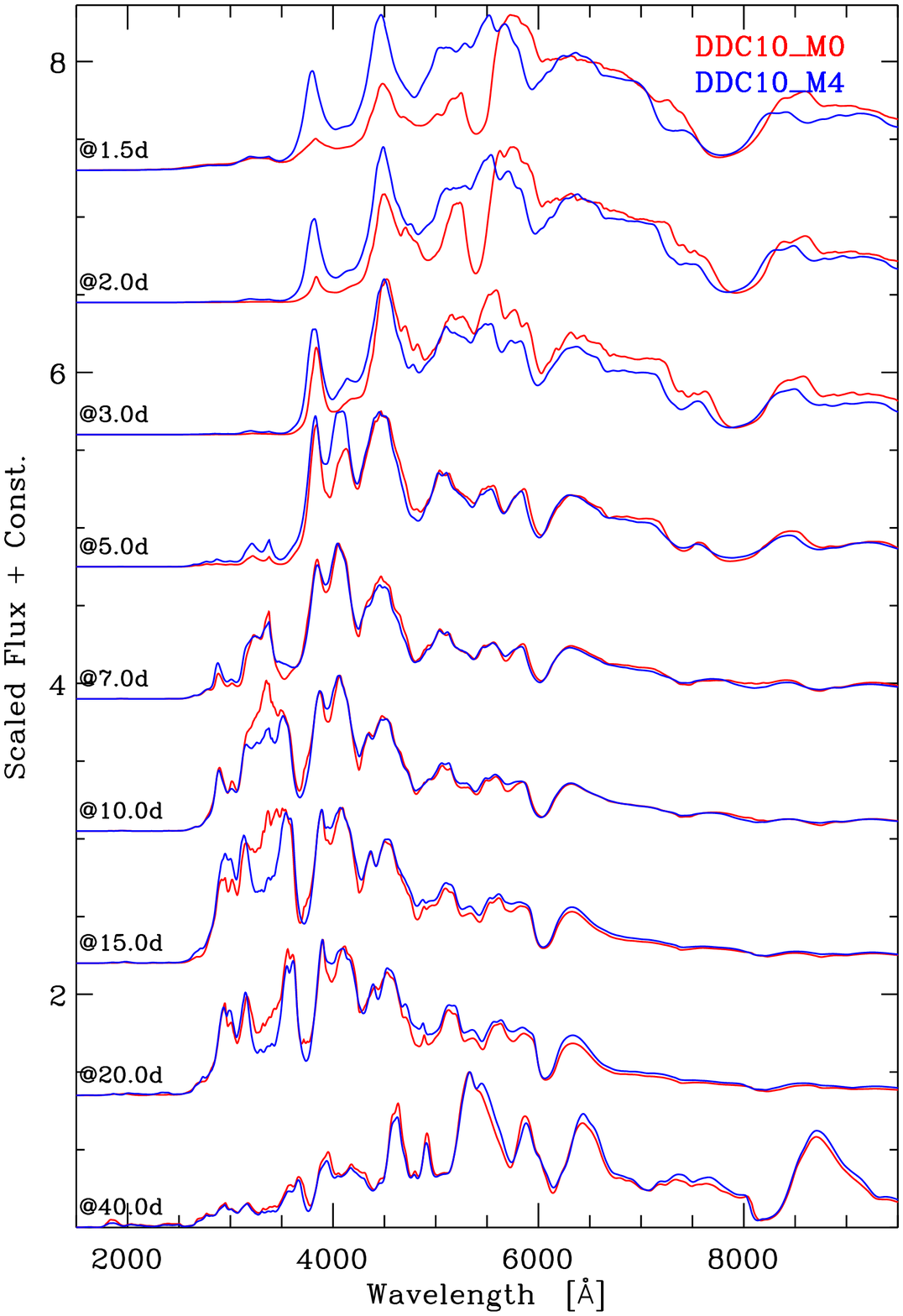,width=8.5cm}
\end{minipage}
\end{flushright}
\caption{{\bf Top left:} Illustration of the contrast in bolometric luminosity between
the reference unmixed model DDC10\_M0 (L$_{\rm bol,ref}$) and variants of the DDC10 model in which
a radial mixing has been applied (models DDC10\_M1 to DDC10\_M4, in order
of increasing mixing).
The strongest deviation is seen at early times, when the SN is very faint.
{\bf Bottom left:} Illustration of the ejecta gas temperature at selected post-explosion times
for the reference model DDC10\_M0 and the highly mixed model DDC10\_M4.
Radial mixing makes the distribution of the \iso{56}Ni much more uniform with depth,
allowing decays and heating in the inner ejecta (where model DDC10\_M0 is
\iso{56}Ni deficient) and above 10000\,\kms\ (where \iso{56}Ni is less abundant).
{\bf Right:} Illustration of the impact on synthetic spectra of allowing for a substantial
radial mixing of the ejecta (DDC10\_M4) or adopting the original 1D chemical
stratification of the hydrodynamical model (DDC10\_M0; this model is also named
DDC10\_A3 in D13 --- see Table~\ref{tab_modset_DDC}).
Apart from early times, when mixing alters both the synthetic spectra
and the bolometric luminosity, the effect of mixing is moderate or even negligible.
}
\label{fig_mix}
\end{figure*}

Because the photosphere (defined as the location where the Rosseland-mean
optical depth integrated inward from the outermost shell is 2/3) 
is located at very large velocities at early times we extend the ejecta grid from
$\sim$\,40000\,\kms\ to 70000\,\kms\ using a linear extrapolation of all fluid
quantities. This may seem very large, but UV photons still interact with the SN ejecta at
velocities $\gtrsim$\,40000\,\kms\ at 0.5\,d.
In model DDC10, the photosphere is, however, at
$\sim$25000\,\kms\ at 0.5\,d, $\sim$\,21000\,\kms\ at 1\,d, and
$\sim$\,18000\,\kms\ at 2\,d after explosion.\footnote{If we were to adopt the flux mean opacity 
the photosphere would be at slightly larger radii} So, the bulk of the radiation emerges from layers that
were optically thick at the time of explosion, at $\sim$\,0.04\,\msun\ below the white dwarf surface, 
and thus well described by the hydrodynamical code. Extrapolation is not ideal but too little is accurately
known about the explosion to make a better guess on the properties of these
outer regions. Overall, the study of SNe Ia at the earliest times is problematic because
the properties will depend  strongly on the initial conditions, at the time of explosion,
which are very poorly known. Further, the outermost regions, which contain very little mass,
 are usually not treated accurately in hydrodynamical simulations.
The inadequacy in the density slope at large velocity is noticeable through the width of the strongest lines, e.g.,
the Ca\two\,8500\,\AA\ triplet, which tend to be much too broad in all our models at early times.
Hence, what is likely to be in error at the earliest times in our calculations is spectral regions influenced by strong line transitions.

The delayed-detonation models (named DDC) presented in \citet{blondin_etal_13},
which cover \nifs\ masses in the range 0.12--0.87\,\msun,
were previously used to compare to radiative properties of SNe Ia at bolometric maximum.
Because these simulations were all initiated at 0.5\,d after explosion,
we can compare their pre-peak properties to those of SN\,2011fe. We find that the explosion models
are extremely faint for 1--2 \,d after explosion and fainter than the pre-discovery brightness
limit for SN 2011fe (Fig.~\ref{fig_slope}). In addition, the models are systematically redder in color and their
brightening rate  in all optical and near-IR bands is steeper than observed for SN\,2011fe  (Fig.~\ref{fig_slope}).
Shifting the observations in time by a day, as done by \citet{mazzali_etal_13}, changes significantly
the magnitude of the offset in brightness and color at such early epochs, although a discrepancy
with observations remains.

\begin{figure*}
\epsfig{file=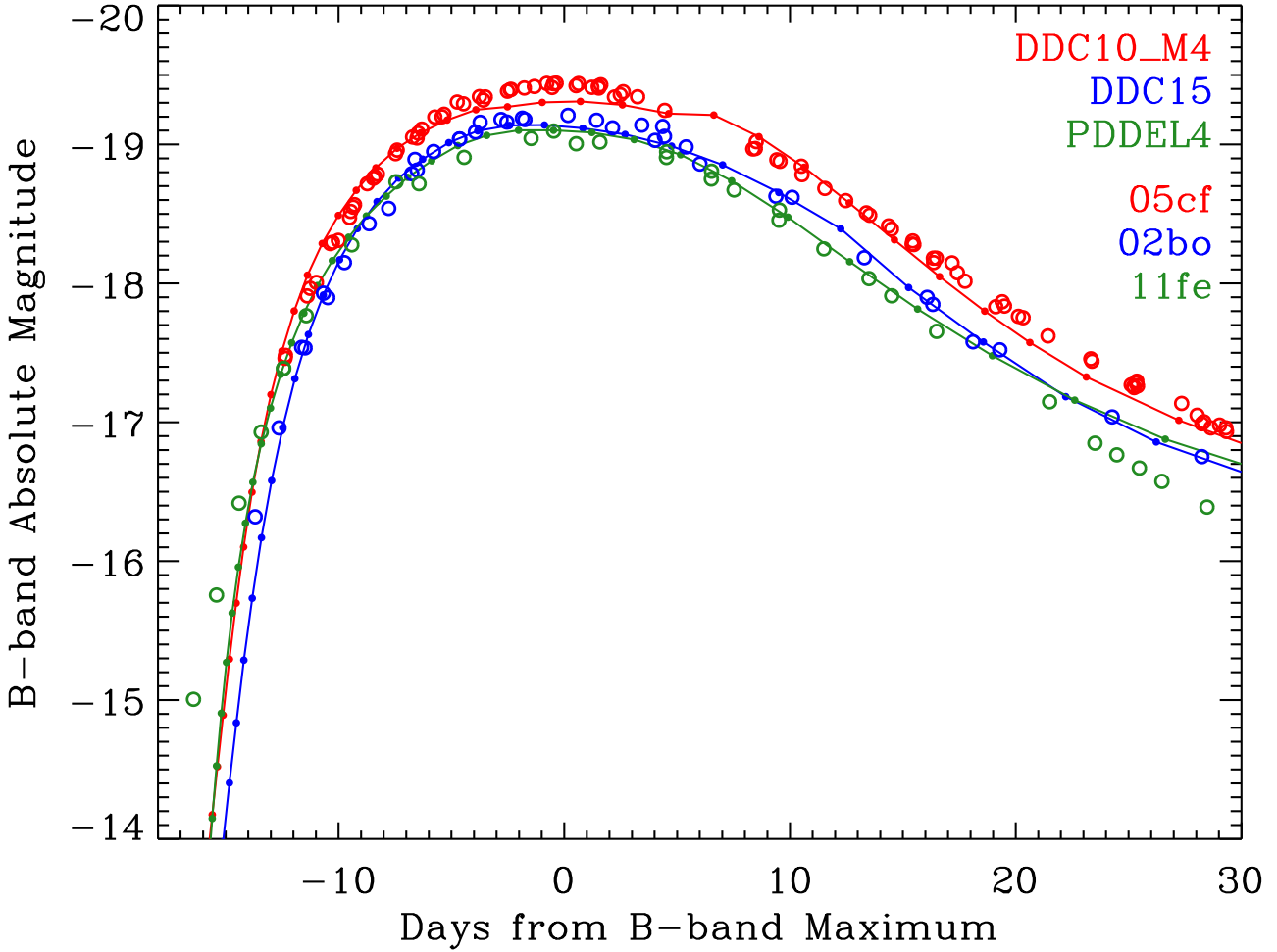,width=17cm}
\caption{
Comparison between the $B$-band light curve of SNe 2005cf (red circles), 2002bo (blue circles),
and 2011fe (green circles), and our \cmfgen\ models DDC10\_M4 (red line), DDC15 (blue line),
and PDDEL4 (green line).
In this figure, we synchronize model and observations at $B$-band maximum, and then, to better
match the light-curve width, we apply a time shift of  $-$0.5, 1.5, and 1\,d to models DDC10\_M4, DDC15,
and PDDEL4, respectively. A detailed study of SN\,2002bo will be presented in Blondin et al. (in preparation).
Details about corrections for distance, extinction, and redshift are given in Section~\ref{sect_obs} and D13.
\label{fig_mag_11fe}
}
\end{figure*}

In all our DDC models, decay energy from \nifs, which causes heating at and below the
photosphere, systematically leads to a hardening of the radiation as the SN brightens.
As a result, none of our delayed-detonation models behaves like a fireball, as observed for SN\,2011fe
\citep{nugent_etal_11}. In contrast, they get considerably bluer as they rise to bolometric
maximum, and the rate of change of that color is greater at earlier times.
This property of our current set of DDC models is robust and holds irrespective of \nifs\ mass.

To investigate the cause of the discrepancies we study which variations in ejecta properties can
alter the early-time color evolution of our delayed-detonation models. We first explore the effects of chemical mixing
(Section~\ref{sect_ddc_mix}). We then consider the influence of treating additional decay routes,
other than the \nifs\ decay chain generally treated (Section~\ref{sect_nuc}).
Some properties of these ``test'' models are given in Table~\ref{tab_modset_DDC}, while the ejecta
properties of DDC models are given in Table~\ref{tab:modinfo}. 

\begin{figure*}
\epsfig{file=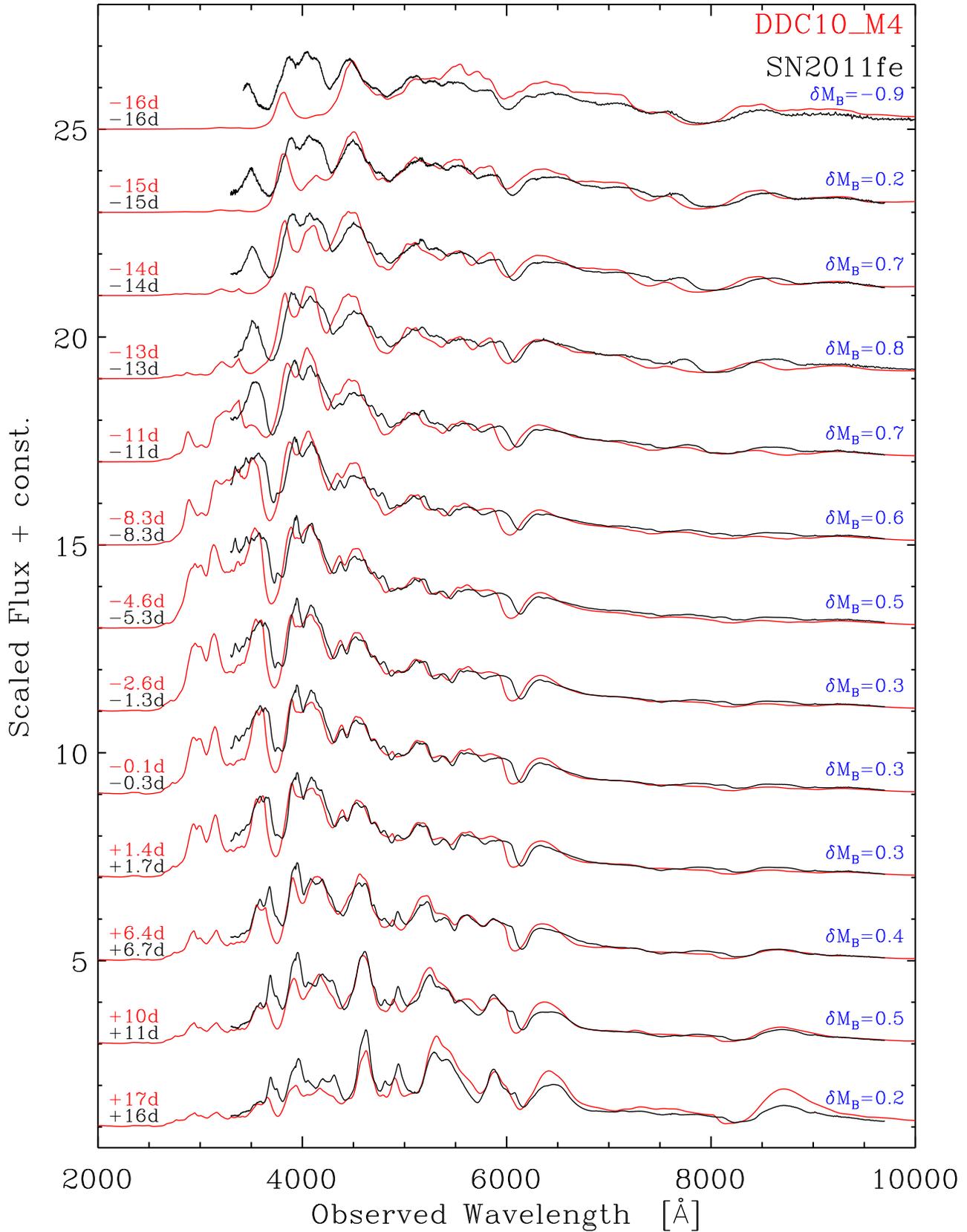,width=17cm}
\caption{
Comparison between model DDC10\,M4 and the observed spectra of SN\,2011fe.
Times are given with respect to $B$-band maximum.
We correct the synthetic flux to account for the distance, redshift, and extinction of SN\,2011fe.
Spectra are scaled vertically to facilitate spectral comparisons, 
although the label on the right gives the true $B$-band
magnitude offset between model and observations at each date.
Despite the strong mixing and the larger \nifs\ mass of 0.623\,\msun\ in model DDC10\_M4
compared to the 0.53\,\msun\ inferred for SN\,2011fe, the model brightness is too low
and its color is to red at the earliest times.
\label{fig_DDC10_M4_spec_11fe}}
\end{figure*}

\subsection{Influence of radial mixing}
\label{sect_ddc_mix}

Delayed detonations in three dimensions reveal the presence of large and small
scale structures, both in the lateral and vertical directions, with significant macroscopic
mixing of fuel and ashes \citep{gamezo_etal_05}. When averaged over angle,
the multi-dimensional structure is reflected primarily through a radial mixing: C and O
are advected inwards to lower velocities, while IMEs and IGEs tend to occupy a much
broader region in velocity space.

As \cmfgen\ is a one-dimensional radiative-transfer code it is not possible to
address directly the multi-dimensional nature of the explosion mechanism, nor
the multi-dimensional effects associated with the radiative transfer. However we
can apply radial mixing, thereby capturing the main feature of multi-dimensional
explosion simulations. This trick has beed used for 1D  radiative-transfer modelling
of core-collapse SNe \citep{blinnikov_etal_00} and, in particular, type Ib SNe
\citep{lucy_91,dessart_etal_12}.

  We test the effect of mixing by running additional sequences in which only the starting
  conditions are modified. Namely, in models DDC10\_MX (X between 1 and 4), the
  chemical structure of model DDC10 is (microscopically) mixed at 100\,s after explosion using
  a characteristic velocity width $v_{\rm mix}$ of 250, 500, 1000, 1500\,\kms\ (Fig.~\ref{fig_init_mix}).
  In practice, we progress from the innermost to the outermost ejecta location and make
  homogeneous all mass shells within a velocity $v_{\rm mix}$ of the local mass shell.\footnote{As we 
  progress outwards, we mix already mixed material. Thus the effect of the adopted algorithm 
  is to smooth over a velocity range much larger than the characteristic velocity width.}
  This softens the composition gradients considerably, but leaves the density structure intact,
   and makes all species present to some extent at all depths for the two models with the highest
   mixing efficiencies.

 In our simplistic approach radial mixing applies to all species.
However, the main effect results from mixing \nifs\ into  high velocity regions
where the original DDC10 model produced little or no \nifs.
In our most extreme model, DDC10\_M4, mixing leads to the
presence of \iso{56}Ni all the way to the inner ejecta shell, causing heating in the regions
below 2000\,\kms\ (the location of the \nifs\ hole in the DDC10\_M0 model).
The contrast with DDC10\_M0 is strong early on 
(these test models have a start time of 1\,d after explosion), but as time
progresses, heat diffuses into the hole in DDC10\_M0 and non-local decay-energy deposition
strengthens, so that the mixed and unmixed models show more comparable temperatures in this region
after bolometric maximum.  

\begin{figure*}
\begin{center}
\epsfig{file=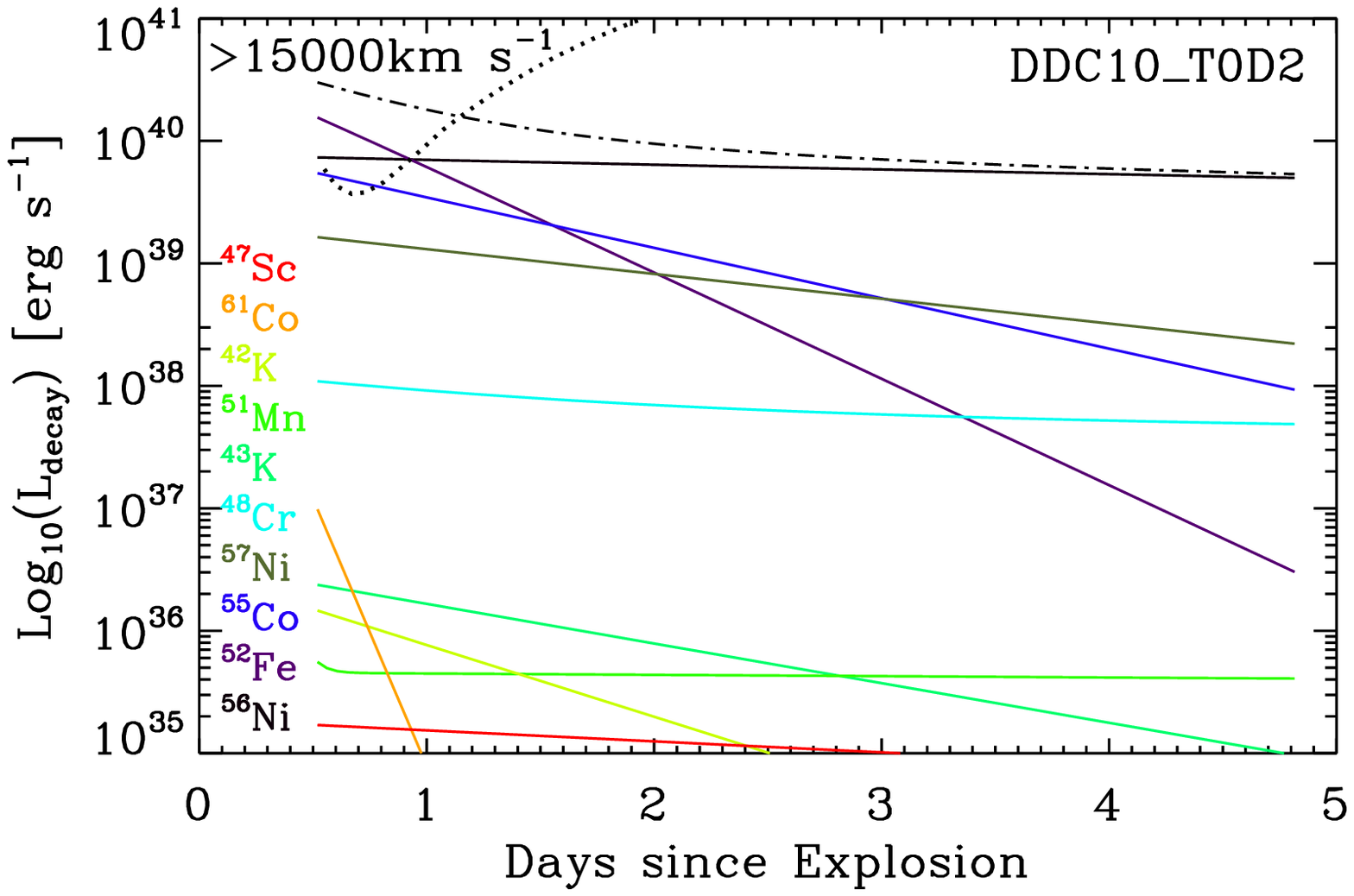,height=5.3cm}
\hspace{2mm}
\epsfig{file=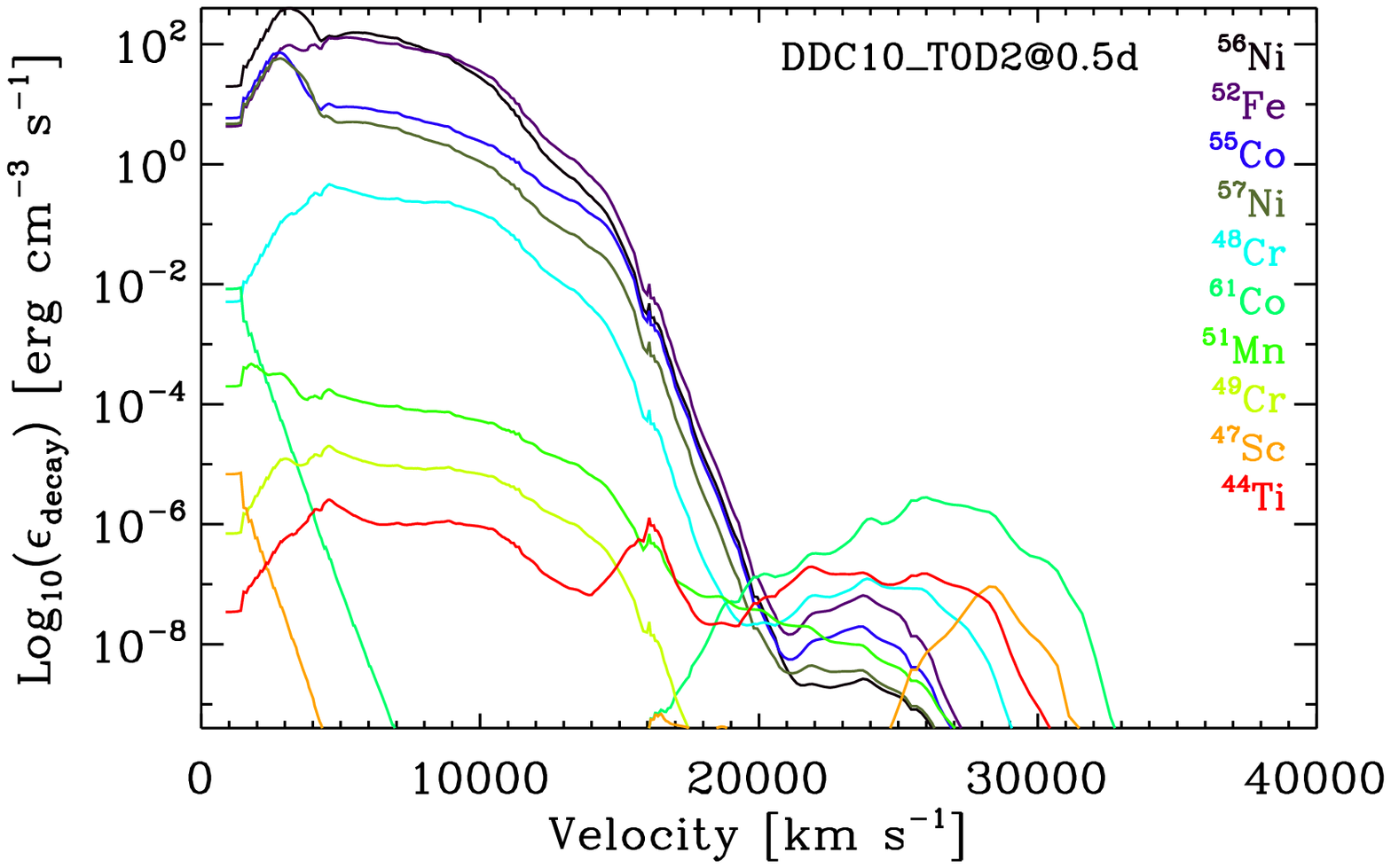,height=5.3cm}
\epsfig{file=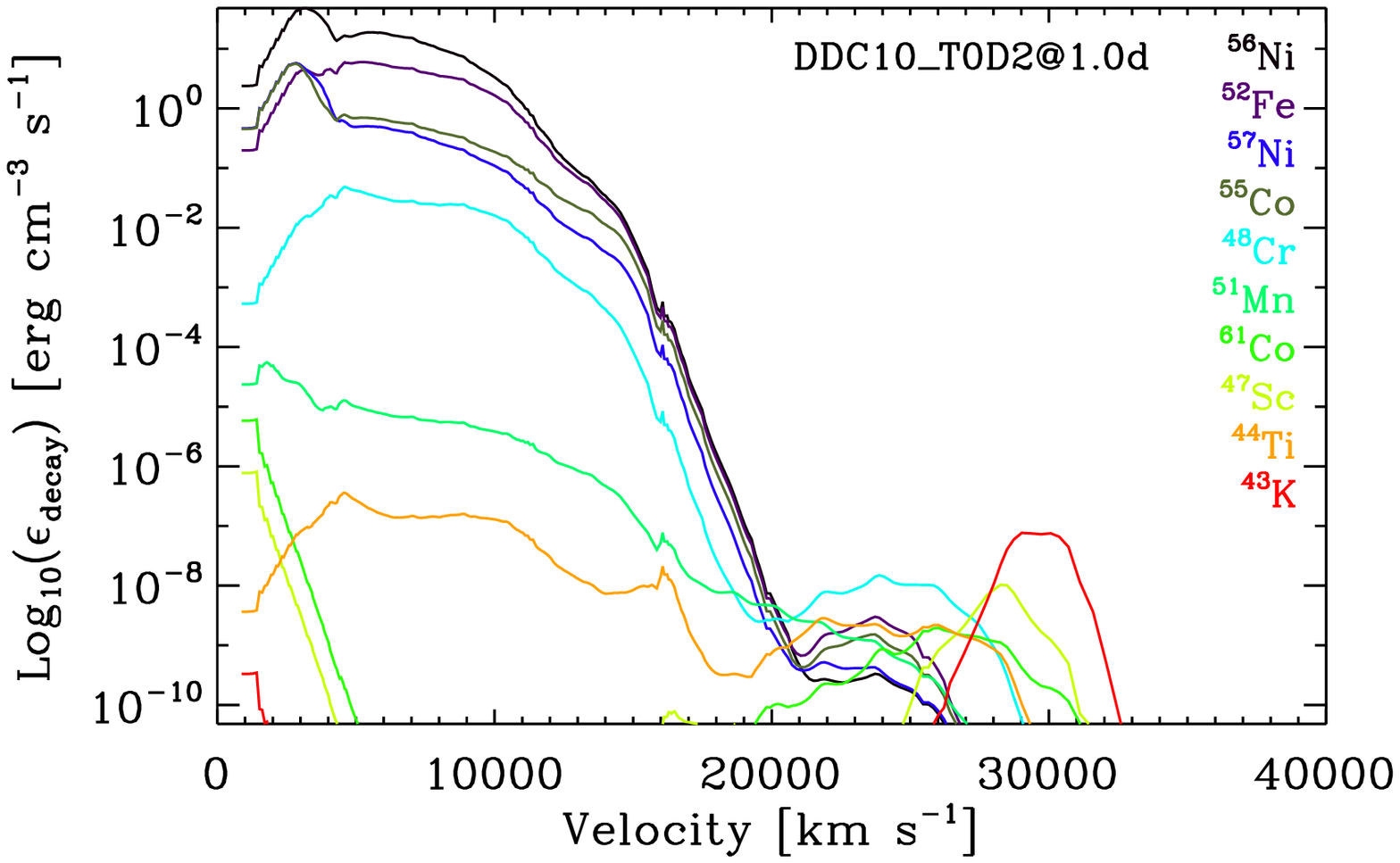,height=5.3cm}
\hspace{2mm}
\epsfig{file=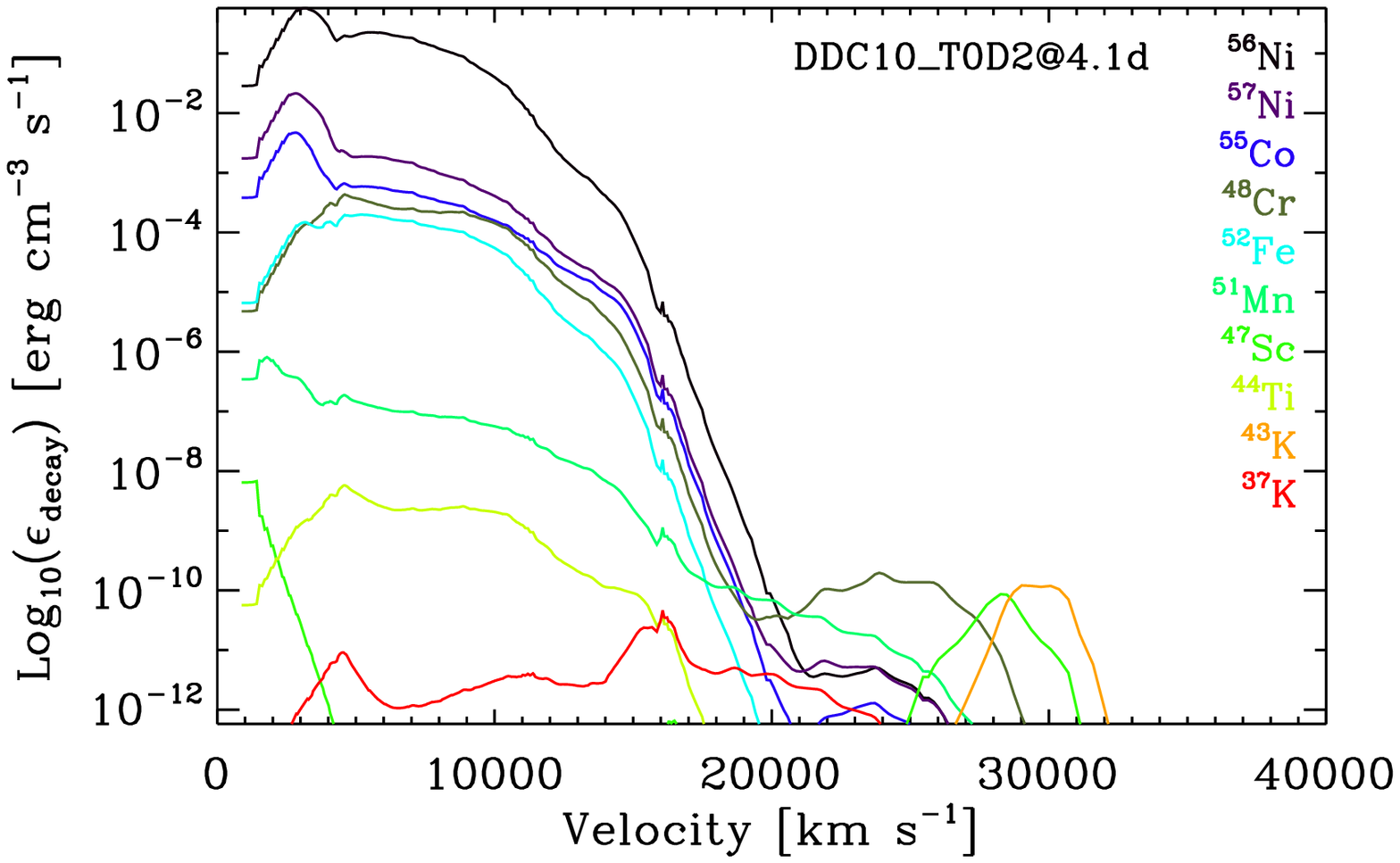,height=5.3cm}
\end{center}
 \caption{{\bf Top left:}
 Illustration of the radioactive-decay power from individual one-step
 and two-step chains for model DDC10\_T0D2.
 In this panel, we include only the contribution from the shells moving faster
 than 15000\,\kms. Among all the chains implemented in \cmfgen, we show only the
 ten strongest power sources.
 The dash-dotted line gives the corresponding total decay energy and the dotted
 line depicts the SN bolometric luminosity computed with \cmfgen\ (epochs during the relaxation
 phase are shown).
 {\bf Other three panels:}
 Illustration of the radioactive-decay emissivity associated with one-step and two-step decay
 chains at 0.5 (top right), 1.0 (bottom left), and 4.1\,d (bottom right). The labels, which refer to each chain plotted, are ordered
 from top to bottom in order of decreasing peak emissivity. Notice the non-negligible contributions
 at high velocity from \iso{61}Co and \iso{43}K.
\label{fig_edecay}
}
\end{figure*}

   Radial mixing influences the bolometric light curve, although the effect is significant
only for the strongest mixing efficiency. The effect is strongest at the earliest times when the 
SN is very faint (e.g., the SN is $\gtrsim$\,1000 times fainter at 1\,d than at bolometric maximum).
and remains visible until a week after explosion  (Fig.~\ref{fig_mix}). This corresponds to epochs when the photosphere
is located at a larger velocity and is also hotter in mixed models. The \iso{56}Ni mass fraction is on the order of 0.01 
at 15000\,\kms\
and effectively zero beyond 20000\,\kms\ in model DDC10\_M0. Mixing causes an enhancement of a factor of a few at
15000\,\kms, and much larger (fractional) enhancements at larger velocities. Although the amount of \nifs\
mass mixed to higher velocities is small it has a very significant effect on the early light curve because
the SN is very faint. 

 In mixed models, the increase in line
blanketing is compensated by the sizable increase in temperature, while the larger outer ejecta optical
depth pushes
the spectrum formation region out to larger velocities. The larger IME mass fraction (from chemical mixing
of IMEs) contributes
to making a broader Si\two\,6355\,\AA\ and enhanced blanketing in some regions, e.g., by Fe\two\
in the $V$ band at 3\,d (Fig.~\ref{fig_mix}).
Beyond bolometric maximum, the SN Ia radiative
properties become insensitive to our adopted mixing. This is easily explained. At advanced times,
all photon ($\gamma$ or optical) mean free paths become large, making the transport of energy non
local. We are therefore not seeing a restricted ejecta shell but instead a large ejecta volume,
with emission biased towards the hottest and densest parts.
These layers are essentially cobalt and iron, and the lack of strong chemical stratification there
makes this region weakly sensitive to mixing.

\begin{figure*}
\centering
\begin{flushleft}
\begin{minipage}{0.5\linewidth}
\epsfig{file=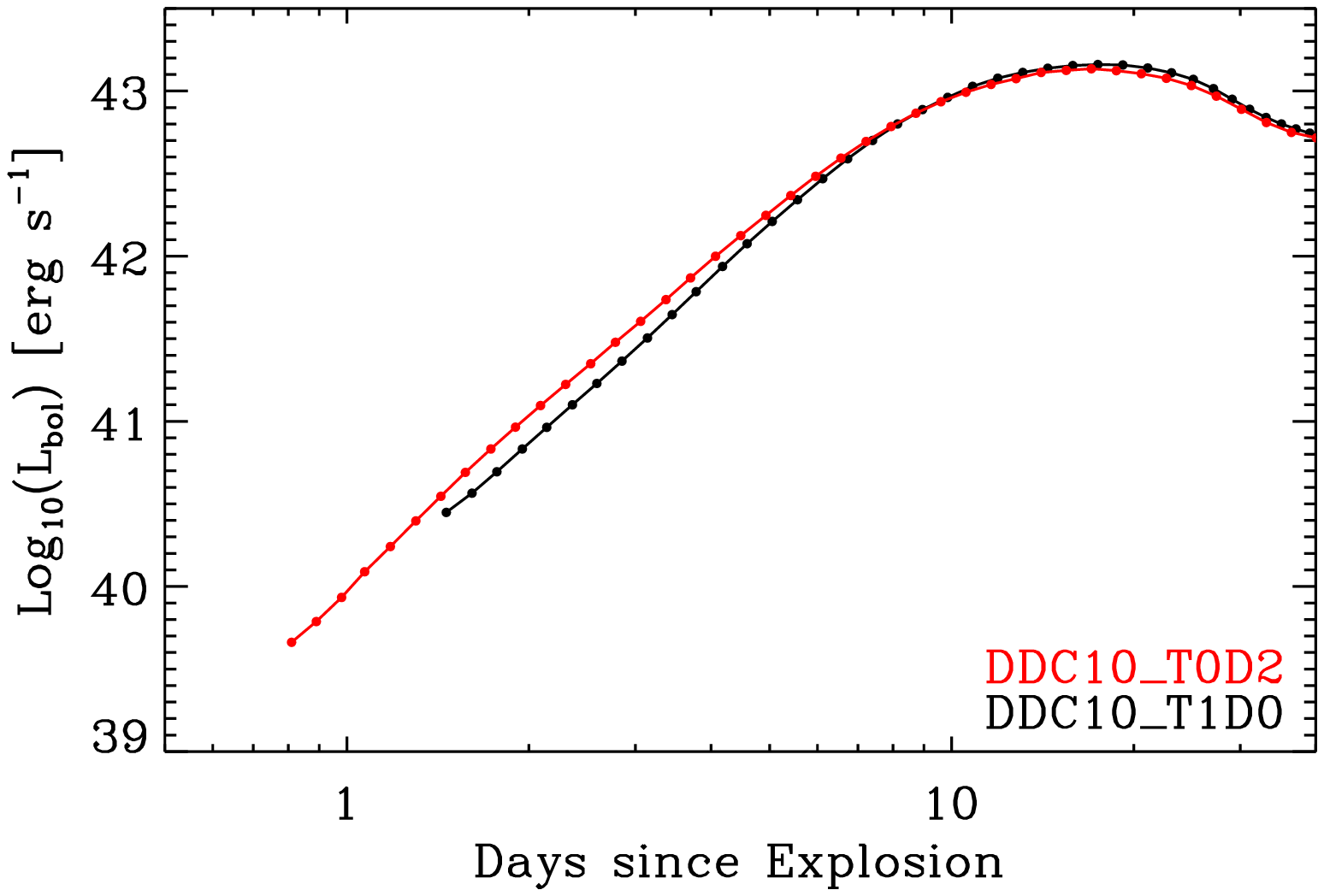,width=8.5cm}
\epsfig{file=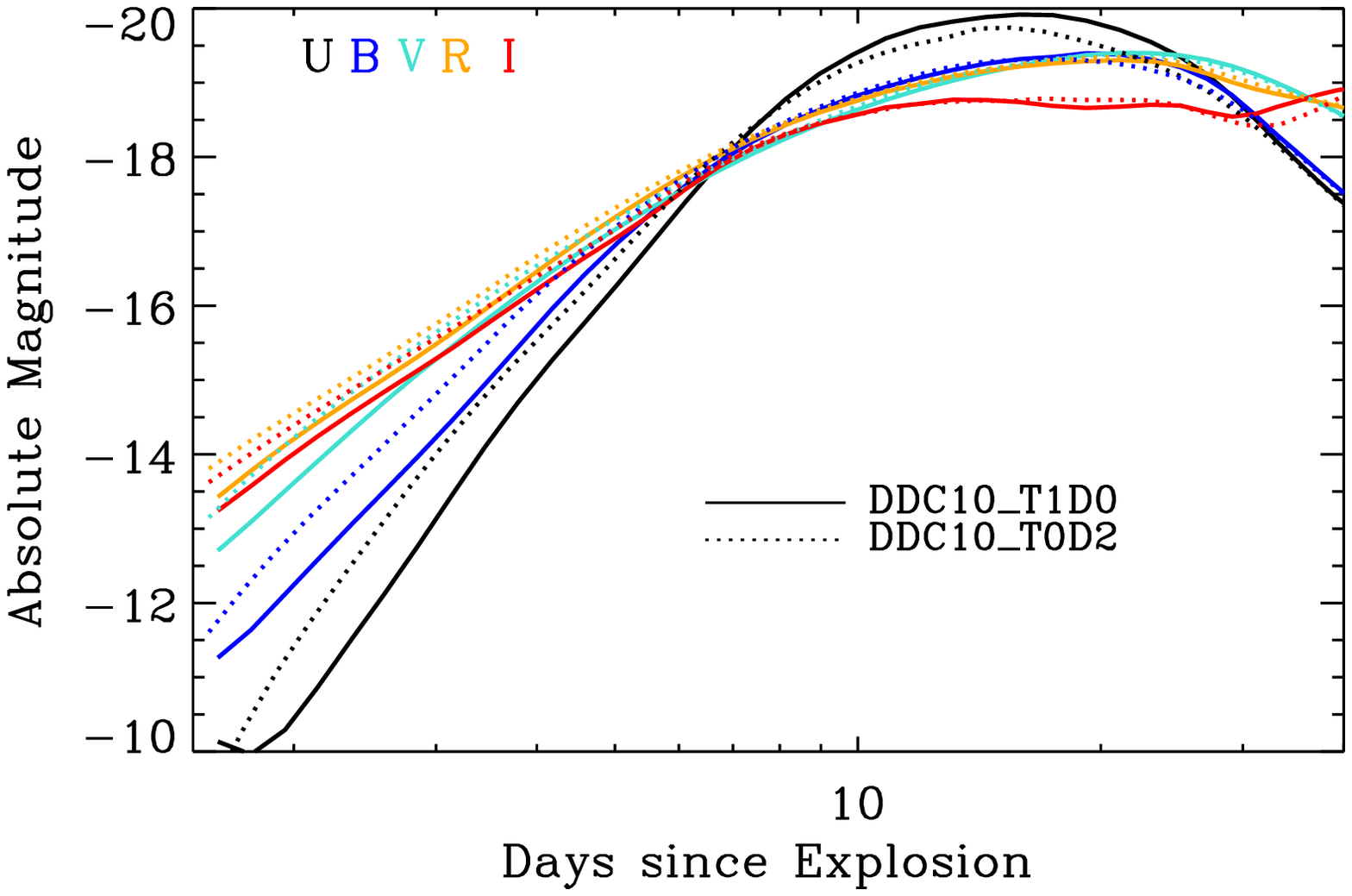,width=8.5cm}
\end{minipage}
\end{flushleft}
\vspace{-11.8cm}
\begin{flushright}
\begin{minipage}{0.5\linewidth}
\epsfig{file=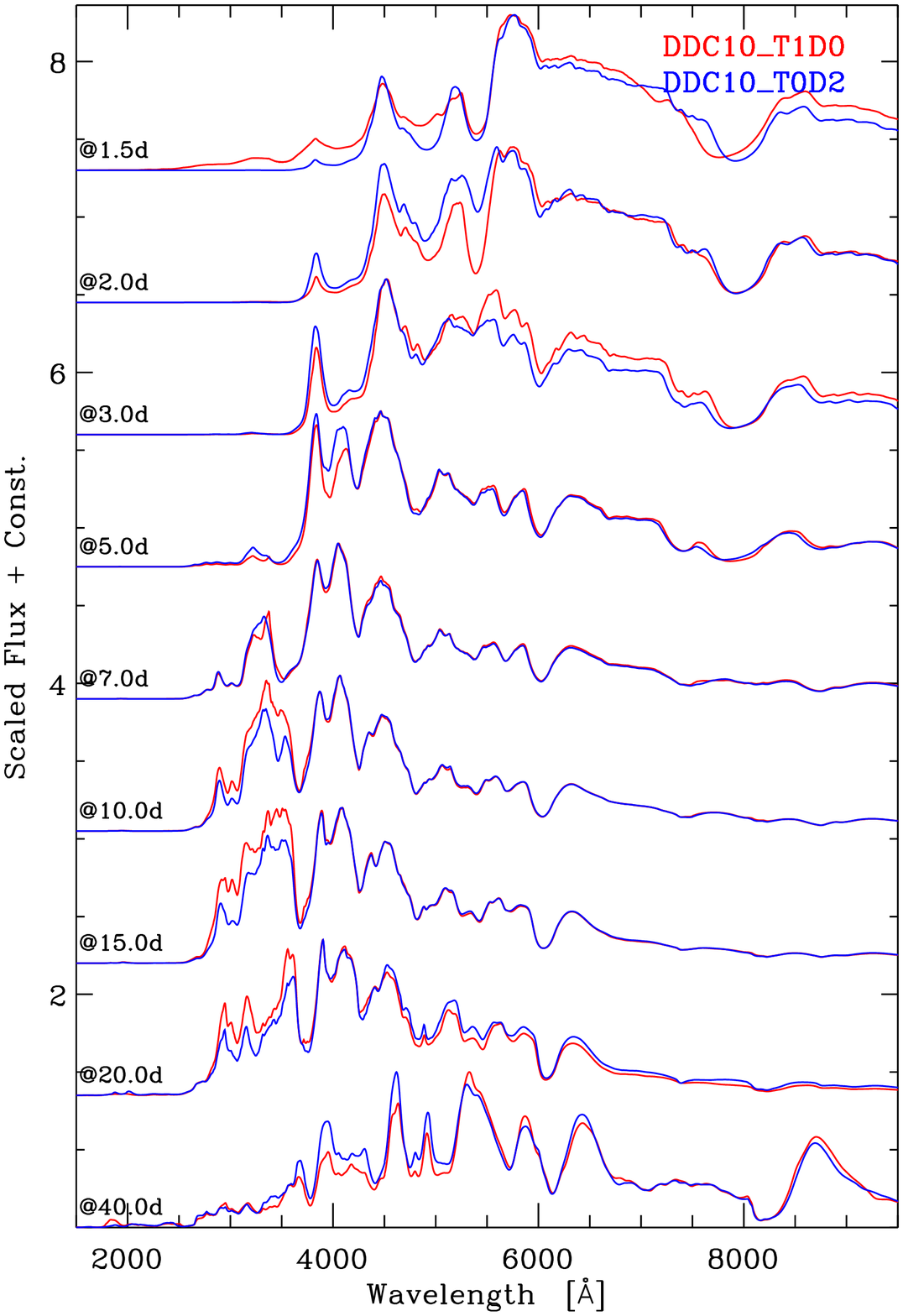,width=8.5cm}
\end{minipage}
\end{flushright}
 \caption{Comparison between DDC10\_T1D0 and DDC10\_T0D2 models (Table~\ref{tab_modset_DDC}).
 The latter uses a start time of 0.5\,d and account for all one-step
and two-step decay chains described in Tables~\ref{tab_nuc1}--\ref{tab_nuc3}.
We show the evolution of the bolometric luminosity (top left), the $UBVRI$ light curves
(bottom left), and the spectral evolution up until 40\,d after explosion (right).
As readily seen from the figures, the inclusion of additional decay chains has too little 
of an effect to reconcile the weakly mixed DDC10 model with the observations (see Fig~\ref{fig_slope}).
 \label{fig_nuc}}
 \end{figure*}

When compared to the $B$-band light curve for three observed SNe (2002bo, 2005cf, and 2011fe), 
model DDC10\_M4 shows a reasonable match to the overall rise to peak (Fig.~\ref{fig_mag_11fe};
see also Fig.~\ref{fig_slope};
additional models are also overlaid and will be discussed in Section~\ref{sect_res}).
However, at the earliest times, the model is significantly fainter than SN\,2011fe, even though
model DDC10\_M4 has 0.65\,\msun\ of \nifs\ compared to 0.53\,\msun\ inferred for SN\,2011fe (this work;
see also, e.g.,  \citealt{pereira_etal_13}). Here, we  have shifted the model by $-0.5$\,d,
and it is clear that a slight mismatch in the inferred explosion time (or in the rise time) translates into a potentially
large magnitude offset at the earliest times, when the SN brightness is predicted to steeply rise.
Even with allowance for a time shift (see, e.g., \citealt{mazzali_etal_13}), SN\,2011fe brightens
along a flatter slope than model DDC10\_M4. SN\,2011fe also has a markedly flatter
brightening slope than SNe 2002bo or 2005cf at comparable pre-maximum epochs.
The recent observations of SN\,2013dy
\citep{zheng_etal_13} further support the existence of a photometric diversity at early times.
Lacking a physical basis in SNe Ia, it is of no surprise  that the fireball model is not universal, but merely
coincidental in SN\,2011fe.

 The relatively bluer colors of model DDC10\_M4 correspond to spectra that are still too red compared
 to early-time observations of SN\,2011fe, although the discrepancy is much reduced compared to
 unmixed models  (Figs.~\ref{fig_slope} and \ref{fig_DDC10_M4_spec_11fe}).
By $\gtrsim$5\,d after explosion, model DDC10\_M4 follows with good fidelity the
 spectral color and morphology, and continues to do so beyond maximum (we show this comparison
 despite the 0.15\,\msun\ difference in \nifs\ mass for the model and this SN). Notable exceptions are
 the Ca\two\ lines and Si\two\,6355\,\AA\, which are systematically broader than observed at all times.
 The velocity structure of this delayed-detonation model is not suitable for
 an event like SN\,2011fe.  Despite these various discrepancies, the fundamental properties of this DDC10\_M4 
 model are in fair agreement with those of SN\,2011fe,
 since no tinkering is applied to the ejecta properties (density, velocity, temperature, composition) throughout
 the time sequence (and the choice of model atoms is the only freedom in the \cmfgen\ simulation; D13).

To conclude, we find that strong mixing has an influence on both the photometric and the spectroscopic
properties of our SN Ia models.
Mixing can cause the SN luminosity to increase by a factor of a few and the color to harden in the optical for 
up to a few days after explosion. 
With strong mixing our DDC10 model is in fair agreement with the observed color of SN\,2011fe.
However, our DDC models tend to show broad
lines, and mixing  tends to exacerbate this property. As a result, mixing would tend to favor the
production of  high-velocity-gradient (HVG) SNe Ia, although this classification tends to be
based on times around and beyond bolometric maximum \citep{benetti_etal_05}. Overall, model DDC10, and
in particular its strongly mixed variants, show broader lines than observed in SN\,2011fe, or SN\,2005cf (see D13).
Mixing can provide a source of SN Ia spectral diversity. For example, it can modulate the trajectory
of absorption maxima at early times in Si\two\,6355\,\AA\ (Fig.~\ref{fig_mix}).

\subsection{Influence of decay chains included}
\label{sect_nuc}

Because of the prevalent role of \iso{56}Ni and \iso{56}Co in controlling SN Ia
radiative properties, the general custom is to include only the decay chain associated with
these two unstable nuclei in SN Ia simulations.
In reality, SN Ia explosions produce a variety of unstable nuclei, either IMEs or IGEs,
present in the inner or outer ejecta, and taking part in two-step or one-step decay chains.
These nuclei show a range of decay lifetimes, from less than a day to years, and can
thus influence SN Ia ejecta on very different time scales.

Here, we improve the consistency of our simulations with a more complete treatment
of decay chains.
The lifetimes, $\gamma$-ray energies and probabilities, and electron-positron emission associated
with all such decays are given in the appendix in Tables~\ref{tab_nuc1}--\ref{tab_nuc3}.
Besides their impact on the internal energy of the gas, these decays modify the composition
and can alter the line-blanketing properties of the ejecta. As discussed in
D13, the decay chain associated with the parent nucleus \iso{48}Cr is
the cause of a 100-fold increase in Ti mass fraction in the spectrum formation
region beyond light curve peak, causing enhanced line blanketing.

\begin{table*}
\footnotesize
\caption{Summary of ejecta properties for the pulsational-delayed-detonation models (PDDEL sequence)
and delayed-detonation models (DDC sequence). 
\label{tab:modinfo}
}
\begin{tabular}{l@{\hspace{1.6mm}}c@{\hspace{1.6mm}}c@{\hspace{1.6mm}}c@{\hspace{1.6mm}}c@{\hspace{1.6mm}}c@{\hspace{1.6mm}}c@{\hspace{1.6mm}}
c@{\hspace{1.6mm}}c@{\hspace{1.6mm}}c@{\hspace{1.6mm}}c@{\hspace{1.6mm}}c@{\hspace{1.6mm}}c@{\hspace{1.6mm}}c@{\hspace{1.6mm}}c@{\hspace{1.6mm}}}
\hline\hline
\multicolumn{1}{c}{Model} & $\rho_{\rm tr}$ & $E_{\rm kin}$ & $v(\nifs)$  & \nifs & Ni & Co & Fe & Ti & Ca & Si & Mg & O & C & $t_B$ \\
 & [\gcc] & [B] & [\kms]  & [M$_{\sun}$] & [M$_{\sun}$] & [M$_{\sun}$] & [M$_{\sun}$] & [M$_{\sun}$] & [M$_{\sun}$] & [M$_{\sun}$] & [M$_{\sun}$] & [M$_{\sun}$] & [M$_{\sun}$] & [d] \\
\hline
PDDEL12        &   1.0(7) & 1.262 &  1.11(4) &      0.253 &      0.268 &   1.88(-2) &      0.101 &   2.37(-5) &   4.56(-2) &      0.489 &   4.21(-3) &      0.103 &   2.11(-2) & 17.24 \\
PDDEL11        &   1.1(7) & 1.236 &  1.14(4) &      0.299 &      0.312 &   2.17(-2) &      0.102 &   2.52(-5) &   4.91(-2) &      0.441 &   2.82(-3) &   7.98(-2) &   2.00(-2) & 16.49 \\
PDDEL9         &   1.3(7) & 1.342 &  1.18(4) &      0.408 &      0.416 &   2.85(-2) &      0.105 &   3.10(-5) &   5.50(-2) &      0.386 &   2.44(-3) &   7.24(-2) &   1.94(-2) & 16.08 \\
PDDEL4         &   1.5(7) & 1.344 &  1.22(4) &      0.529 &      0.530 &   3.58(-2) &      0.108 &   3.43(-5) &   5.40(-2) &      0.307 &   1.64(-3) &   4.70(-2) &   1.85(-2) & 16.60 \\
PDDEL7         &   1.6(7) & 1.336 &  1.25(4) &      0.604 &      0.602 &   4.02(-2) &      0.107 &   3.07(-5) &   5.03(-2) &      0.258 &   1.52(-3) &   4.34(-2) &   1.79(-2) & 17.65 \\
PDDEL3         &   1.8(7) & 1.353 &  1.26(4) &      0.685 &      0.680 &   4.51(-2) &      0.107 &   3.01(-5) &   4.64(-2) &      0.218 &   1.46(-3) &   4.04(-2) &   1.77(-2) & 18.21 \\
PDDEL1         &   2.0(7) & 1.398 &  1.28(4) &      0.758 &      0.751 &   4.95(-2) &      0.107 &   2.99(-5) &   4.23(-2) &      0.190 &   1.48(-3) &   3.99(-2) &   1.73(-2) & 18.18 \\
\hline
\multicolumn{1}{c}{Model} & $\rho_{\rm tr}$ & $E_{\rm kin}$ & $v(\nifs)$  & \nifs & Ni & Co & Fe & Ti & Ca & Si & Mg & O & C & $t_B$ \\
 & [\gcc] & [B] & [\kms]  & [M$_{\sun}$] & [M$_{\sun}$] & [M$_{\sun}$] & [M$_{\sun}$] & [M$_{\sun}$] & [M$_{\sun}$] & [M$_{\sun}$] & [M$_{\sun}$] & [M$_{\sun}$] & [M$_{\sun}$] & [d] \\
\hline
DDC25          &   8.0(6) & 1.185 &  8.49(3) &      0.119 &      0.142 &   9.69(-3) &   9.80(-2) &   1.13(-4) &   2.41(-2) &      0.485 &   3.72(-2) &      0.283 &   2.16(-2) & 19.82 \\
DDC22          &   1.1(7) & 1.345 &  9.80(3) &      0.211 &      0.231 &   1.59(-2) &      0.107 &   1.17(-4) &   4.15(-2) &      0.483 &   2.57(-2) &      0.209 &   8.26(-3) & 17.49 \\
DDC20          &   1.3(7) & 1.442 &  1.03(4) &      0.300 &      0.315 &   2.15(-2) &      0.110 &   1.12(-4) &   4.72(-2) &      0.426 &   2.10(-2) &      0.170 &   5.12(-3) & 17.21 \\
DDC17          &   1.6(7) & 1.459 &  1.08(4) &      0.412 &      0.421 &   2.84(-2) &      0.112 &   1.14(-4) &   4.73(-2) &      0.353 &   1.79(-2) &      0.152 &   3.80(-3) & 17.24 \\
DDC15          &   1.8(7) & 1.465 &  1.12(4) &      0.511 &      0.516 &   3.44(-2) &      0.114 &   1.11(-4) &   4.53(-2) &      0.306 &   1.14(-2) &      0.105 &   2.73(-3) & 17.09 \\
DDC10          &   2.3(7) & 1.520 &  1.16(4) &      0.623 &      0.622 &   4.11(-2) &      0.115 &   1.10(-4) &   4.10(-2) &      0.257 &   9.95(-3) &      0.101 &   2.16(-3) & 17.69 \\
DDC6           &   2.7(7) & 1.530 &  1.20(4) &      0.722 &      0.718 &   4.72(-2) &      0.116 &   1.07(-4) &   3.52(-2) &      0.216 &   7.28(-3) &   8.35(-2) &   1.81(-3) & 18.03 \\
DDC0           &   3.5(7) & 1.573 &  1.29(4) &      0.869 &      0.872 &   5.58(-2) &      0.102 &   1.17(-4) &   2.49(-2) &      0.160 &   3.58(-3) &   5.18(-2) &   1.20(-3) & 17.67 \\
\hline
\end{tabular}
\flushleft
{\bf Notes:}
Numbers in parenthesis correspond to powers of ten.
$\rho_{\rm tr}$ is the transition density at which the deflagration is artificially turned into a detonation;
$E_{\rm kin}$ is the asymptotic kinetic energy (units: $1\mathrm{B} \equiv 1 \mathrm{Bethe} = 10^{51}$\,erg);
$v(\nifs)$ is the velocity of the ejecta shell that bounds 99\% of the total \nifs\ mass.
The \iso{56}Ni mass is given at $t\approx$\,0\,s, while the other cumulative masses correspond to a time of 0.5\,d after explosion.
$t_B$ is the $B$-band rise time. All these models were mixed using $v_{\rm mix}=$\,400\,\kms.
\end{table*}

We thus run some variants of model DDC10 in which different sets of decay chains
are treated, corresponding to simulations with suffix ``D''.
Simulations that include the \nifs\ two-step decay chain
only are named ``D0''. Simulations that include all two-step decay chains
have suffix ``D1'' (see D13). Simulations that include all fifteen decay chains
(six one-step chains and nine two-step chains) have suffix ``D2''.
 When suffix ``T'' is specified, ``T0'' refers to
time sequences started with \cmfgen\ at 0.5\,d and suffix ``T1'' refers to
sequences started at 1\,d after explosion.

At early post-explosion times, we find that numerous
chains contribute more energy than that of \nifs, in particular in the outer ejecta,
above 15000-20000\,\kms\ where \nifs\ is under-abundant (Fig.~\ref{fig_edecay}).
We find that the chains
associated with \iso{42}K, \iso{43}K, \iso{61}Co, and \iso{48}Cr dominate the
decay energy above 20000\,\kms\ at 1\,d.
By 4\,d after explosion, the same regions are primarily influenced by the
decay of \iso{43}K, \iso{48}Cr, and \iso{47}Sc. At smaller velocities, besides
\nifs, we find \iso{52}Fe, \iso{55}Co, and \iso{57}Ni.
Despite the treatment of these additional contributions in \cmfgen, the
resulting synthetic spectra change little, and thus remain quite red
(Fig.~\ref{fig_nuc}). The main effect is a global brightening in all bands with
a slight hardening of the spectrum.
For example, for model DDC10\_T0D2 compared to model DDC10\_T1D0. 
$B-R$ goes from 2.0 to 1.7\,mag at 2\,d, which corresponds to a 0.3\,mag shift only.
At bolometric maximum, the $U$-band flux offset is caused by
enhanced blanketing from Ti, whose mass fraction is increased 100-fold
in model  DDC10\_T0D2 through the decay of \iso{48}Cr (D13).
Overall, extra decays induce too small a correction to the
DDC10 model to bring it into agreement with the observations of SN\,2011fe.

Up to $\sim$\,10\,d after explosion, we assume local energy deposition because
the bulk of radioactive decays occurs at high optical depth where the $\gamma$-ray mean
free path is small compared to the size of the SN ejecta. Indeed, allowing 
for non-local energy deposition earlier on in \cmfgen\ causes only minute alterations to the
line profile morphology and no change in color or luminosity in our models.
SN\,2011fe, which is more luminous and bluer that any of our DDC models
at 1-2\,d after explosion (Fig.~\ref{fig_slope}), is typically sub-luminous at peak \citep{roepke_etal_12},
so the \nifs\ should be buried somewhat deep in the ejecta, hence unlikely to affect drastically the outer
ejecta layers at $\sim$\,1\,d.
{\it One is thus led to question whether $^{\it 56}$Ni is the only quantity
controlling the SN Ia radiation properties at the earliest times.}

\section{Pulsational-delayed detonation models of SN\lowercase{e} I\lowercase{a}}
\label{sect_setup}

In this section we discuss the potential merits of the pulsational-delayed detonation model
to explain a number of SN Ia properties, in particular those that we do not reproduce
satisfactorily with the ``standard" delayed-detonation model.\footnote{It is unfortunately
not possible at present to discuss the ``universal" radiative properties of a given explosion model
because many differences are in fact tied to the radiative-transfer treatment. For example, \citet{sim_etal_13}
find fundamental problems with the ``standard" delayed detonation model of SNe Ia, while for similar
models, D13 find a promising agreement with observations
suggestive of the broad adequacy of that explosion configuration.}
This explosion configuration in a Chandrasekhar mass white dwarf has been previously studied in
\citet{hoeflich_etal_96} and confronted with observed SN Ia light curves.

\begin{figure*}
\epsfig{file=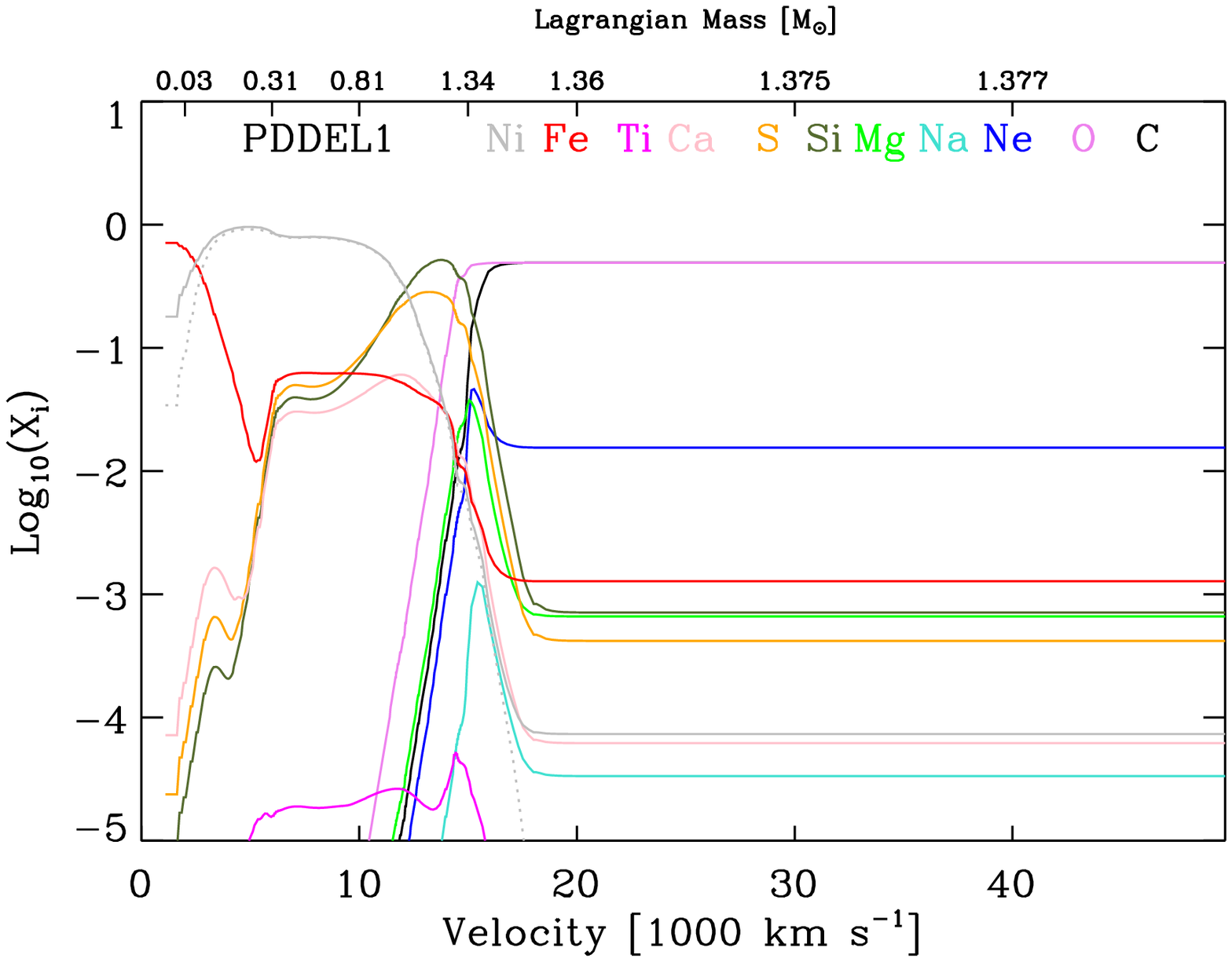,width=8.5cm}
\hspace{2mm}
\epsfig{file=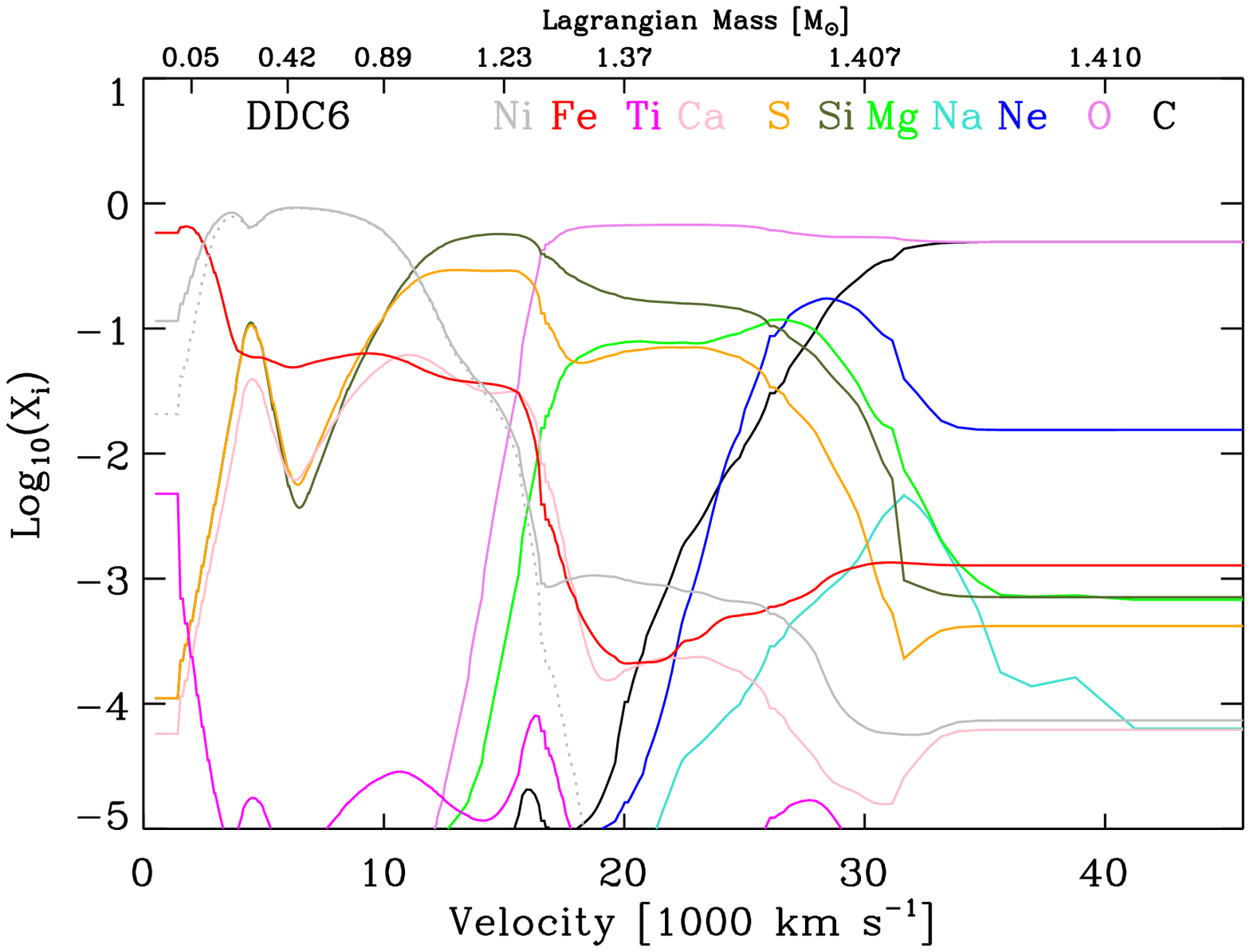,width=8.5cm}
\caption{Chemical composition versus velocity and mass (top axis) for models PDDEL1 (left)
and DDC6 (right). While they have very similar cumulative chemical yields, their distribution is different.
Model PDDEL1 shows a stronger confinement of chemical species, i.e.,
species tend to be present over narrower velocity ranges.
This contrast in chemical stratification holds between all PDDEL models and their DDC counterpart
(see Table~\ref{tab:modinfo}).
\label{fig_comp}
}
\end{figure*}

For this work, we produce pulsational-delayed detonations somewhat artificially.
We first initiate a deflagration in a Chandrasekhar-mass white dwarf.
After a modest expansion of the white dwarf, we stop nuclear burning.
Upon the subsequent infall and compression of the white dwarf material
that is still bound, a detonation is initiated at a prescribed transition density, whose
value is tuned to produce ejecta with a different \nifs\ mass.
Importantly, the material that becomes loosely bound during the first deflagration expands
sufficiently to avoid burning by the subsequent detonation. This allows more carbon to
remain unburnt compared to the ``standard" delayed-detonation scenario, which generally leaves
no unburnt material at all \citep{K91a}.

\begin{figure*}
\epsfig{file=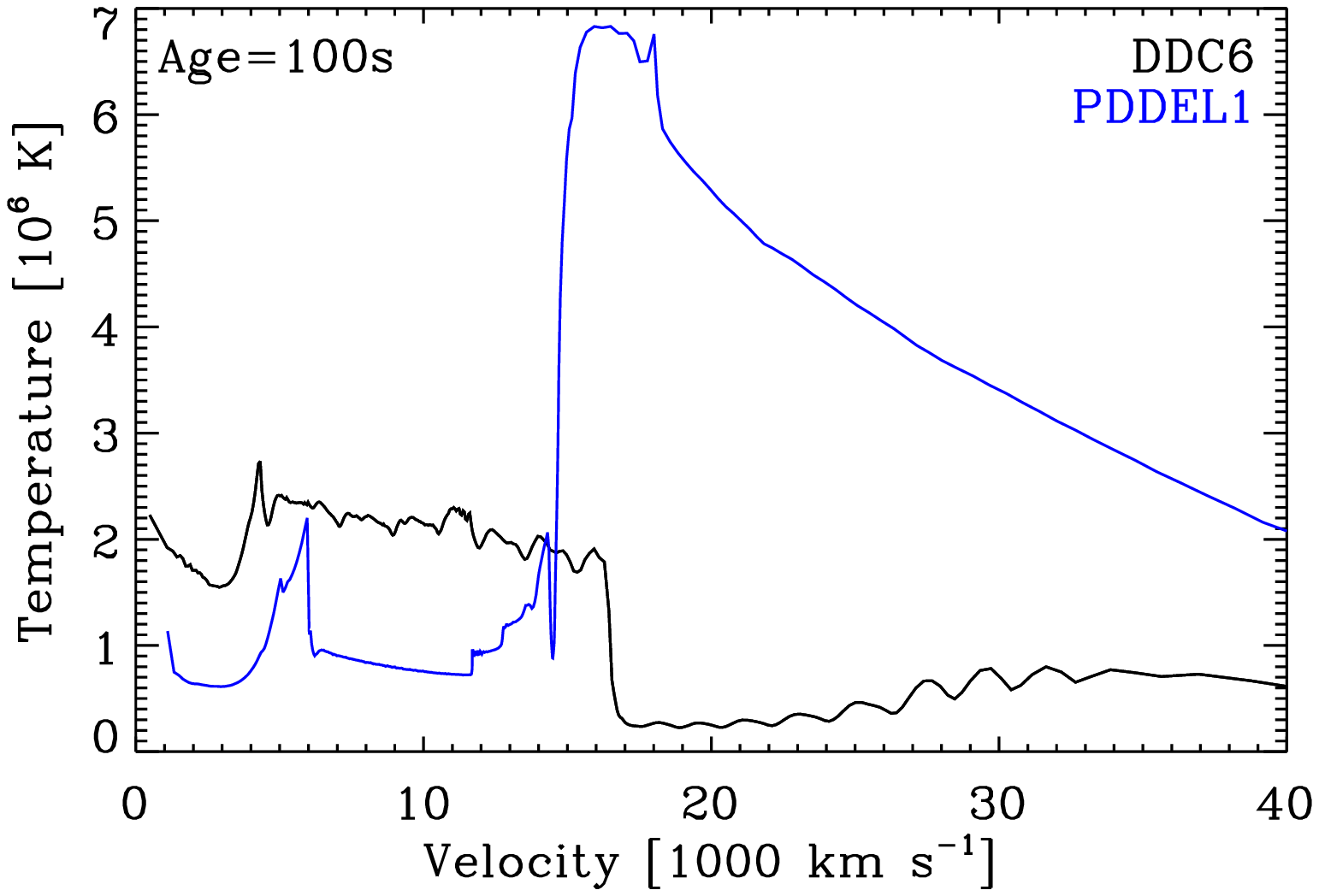,height=5.3cm}
\hspace{2mm}
\epsfig{file=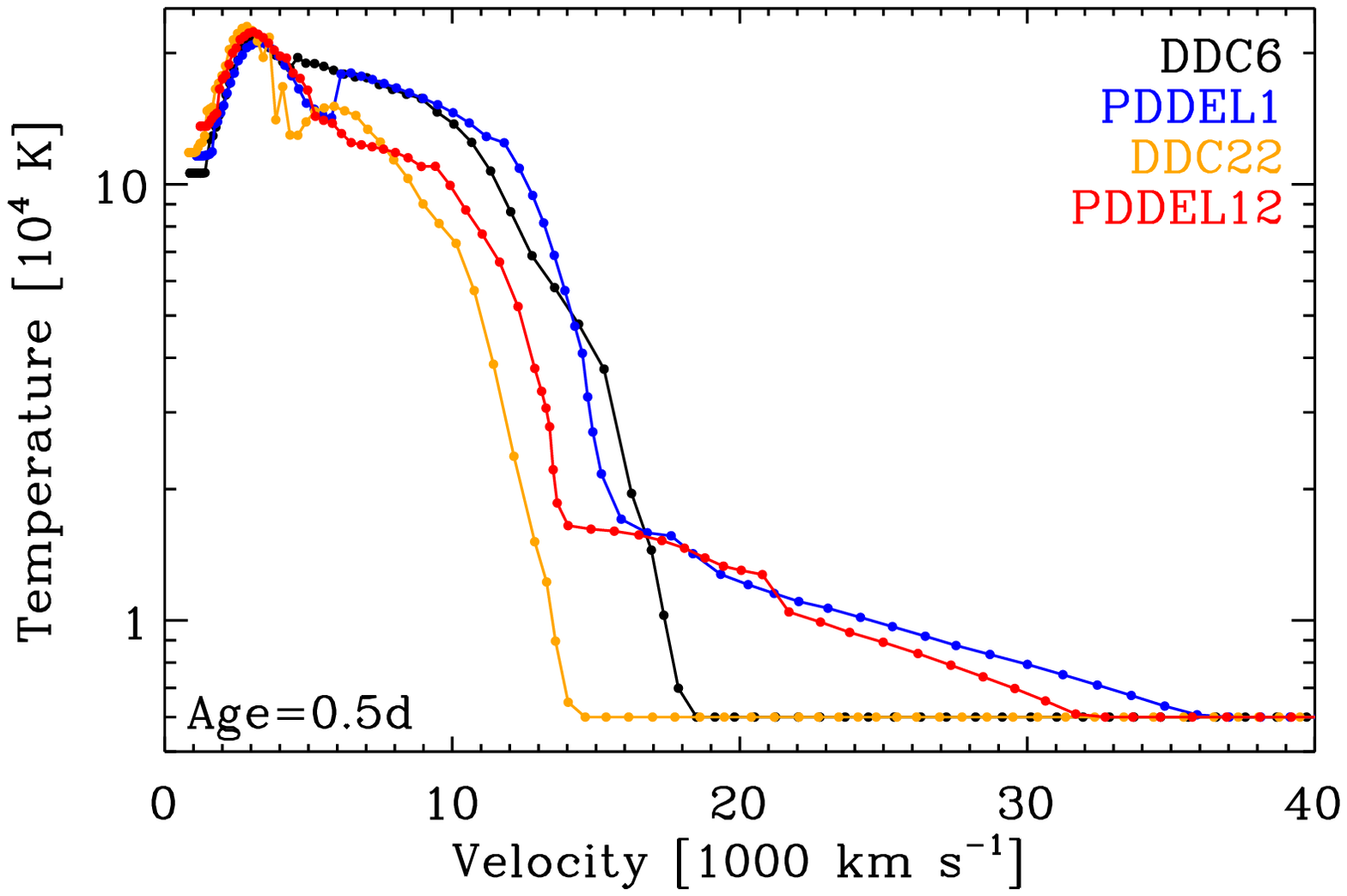,height=5.3cm}
\caption{{\bf Left:} Comparison of the ejecta temperature structure from models DDC6 (``standard'' delayed detonation)
and PDDEL1 (pulsational-delayed detonation) at 100\,s after explosion.
At this time, the influence of \nifs\ decay heating is negligible. The temperature structure
is instead set by the hydrodynamics of the explosion, the energy release from combustion, and cooling from
expansion. In the PDDEL model, the hydrodynamic interaction of the inner and outer white dwarf layers is the origin of
a large offset in temperature compared to the DDC counterpart.
{\bf Right:}
Comparison of the ejecta temperature structure for two sets of DDC and PDDEL models at large and low
\nifs\ mass at 0.5\,d after explosion. At this time, \nifs-decay heating controls the temperature in the
inner ejecta but plays a negligible role beyond $\sim$\,15000\,\kms.
The outer temperature structure is thus strongly influenced by the original temperature, which is much larger
in  pulsational-delayed detonation models. Symbols give the location of the \cmfgen\ grid points
for the first time step.  
\label{fig_comp_tinit}
}
\end{figure*}

Pulsational-delayed-detonation models correspond to the  PDDEL sequence,
which is composed of individual models named PDDELm with
m being 1, 3, 7, 4, 9, 11, and 12,  in order of decreasing \nifs\ mass, from 0.76 to 0.25\,\msun.
Each model retains about $\lesssim$\,0.02\,\msun\ of unburnt carbon, irrespective of \nifs\ mass --- the rest
of the ejecta composition is typical of ``standard" delayed-detonation models.
Standard delayed-detonation models correspond to the DDC sequence (used here for comparison).
It is composed of models whose properties at bolometric maximum are
presented in \citet{blondin_etal_13}. Individual model names are DDCn with n being 1, 6, 10, 15, 17,
20, 22, and 25, in order of decreasing \nifs\ mass, from 0.87 to 0.12\,\msun.
The mass of unburnt carbon in DDC models is typically a factor of 2--10 lower than in PDDELm models.
We summarize the chemical yields of our pulsational-delayed-detonation
models as well as those for the ``standard" delayed-detonation models in Table~\ref{tab:modinfo}.
For all DDC and PDDEL simulations discussed in the next section, we apply the same mixing procedure
(we use $v_{\rm mix}=$\,400\,\kms)
and account for the same set of decay chains --- all two-step and one-step decay chains
presented in Table~\ref{tab_nuc1}--\ref{tab_nuc3} are included in the calculation.
Hence, our simulations differ only in initial ejecta properties.

The ejecta structures of models PDDEL are significantly different from those of DDC models, forming two distinct families.
This stems primarily from the hydrodynamical interaction that takes place in PDDEL models, between
the outer infalling white dwarf and the inner layers of the white dwarf where the detonation
goes off. This reduces the amount of mass at large velocity.
It also produces a steep change in density
(hereafter termed the cliff) at the lagrangian mass where the interaction takes place and a
much larger temperature of the lower density shocked material than in the ``standard" delayed-detonation scenario.
The density cliff is located at slightly larger velocities in models with a larger \nifs\ mass,
but the scatter is small, with a mean velocity of $\sim$\,14000\,\kms\ (in practice, this value can be altered
by slight adjustments in the numerical procedure, so it is not obtained robustly from first principles).

\begin{figure}
\epsfig{file=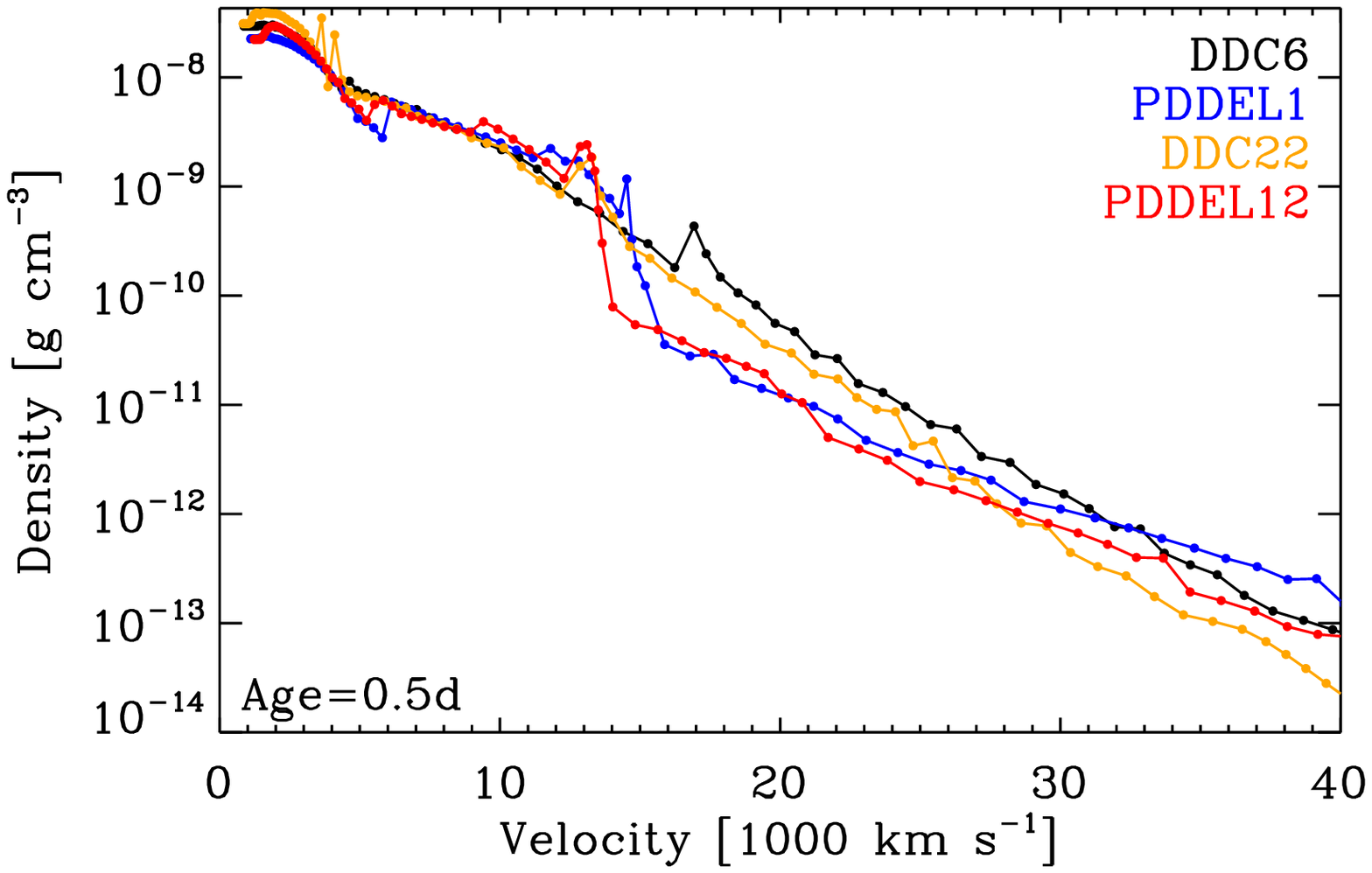,height=5.3cm}
\caption{Same as right panel of Fig~\ref{fig_comp_tinit}, but now for the mass density at 0.5\,d.
Note the presence of a sizable density cliff in PDDEL models.
\label{fig_comp_dinit}
}
\end{figure}

To illustrate differences, we use from each DDC and PDDEL set a couple of models with comparable
\nifs\ mass, at a high and at a low value. For the set with a large \nifs\ mass, we use models
DDC6 and PDDEL1 (initially with 0.72 and 0.76\,\msun\ of \nifs, respectively).
For the set with a low \nifs\ mass, we use models DDC22 and PDDEL12 (initially with 0.21 and 0.25\,\msun, respectively).
Figure~\ref{fig_comp} illustrates the different stratification in {\it velocity} space for the ejecta of models
DDC6 and PDDEL1. Although they synthesize a comparable mass of \nifs\ and Si, the Mg/Ti/O/C yields
differ by a factor of 7/5/2/0.1 (Mg, Ti, O, and C are under-abundant species relative to \nifs\ and Si).
These non-trivial variations could be a clear source of diversity at a given \nifs\ mass (or peak luminosity).
Models of the PDDEL sequence also systematically show a stronger confinement of chemical species.
In particular, nuclear processed IMEs are restricted to a narrower velocity range than
in ``standard" delayed-detonation models.

\begin{figure*}
\epsfig{file=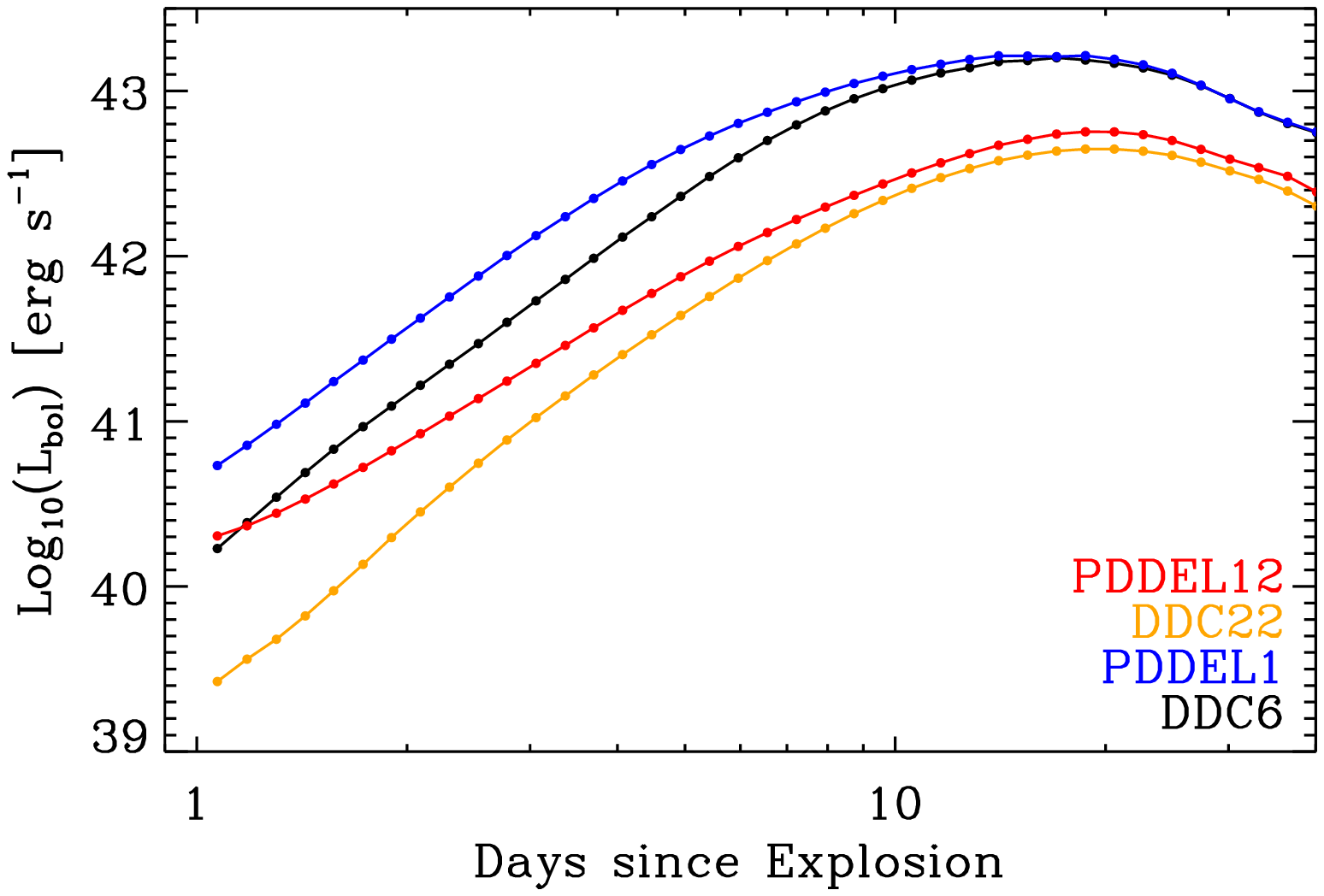,height=5.3cm}
\hspace{2mm}
\epsfig{file=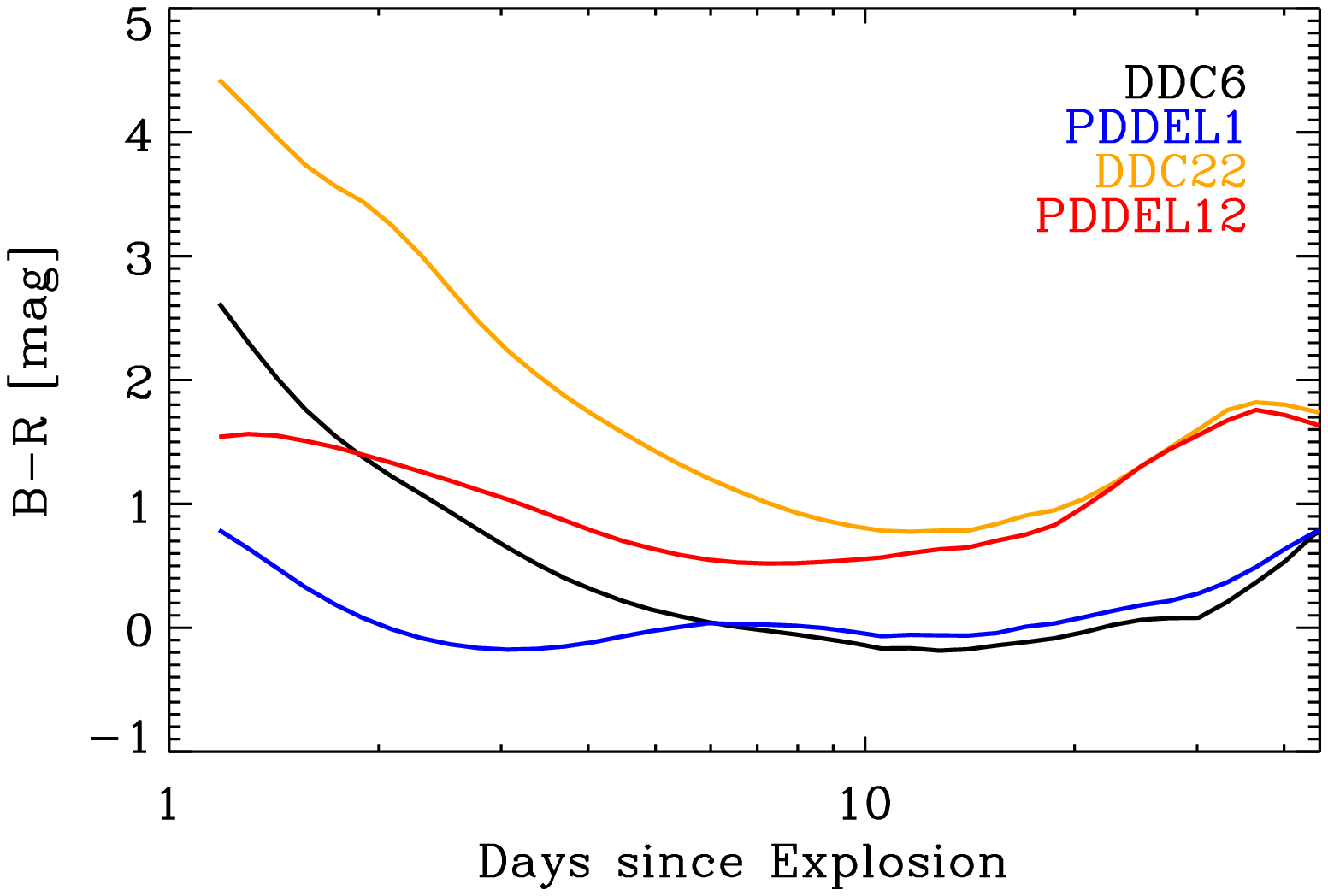,height=5.3cm}
\caption{{\bf Left:} Comparison between bolometric light curves of DDC and PDDEL models.
The peak luminosity is nearly identical within each pair, because of the comparable \nifs\ mass
of associated models. In contrast, the early time luminosity is much higher in PDDEL than in DDC models,
with a weaker dependence on \nifs\ mass.
{\bf Right:} Evolution of the $B-R$ color for DDC and PDDEL models. DDC models undergo a strong
shift of their color to the blue as they evolve up to bolometric maximum.
In contrast, PDDEL models start bluer, even if endowed with a low \nifs\ mass, and undergo only a modest
color evolution on the way to bolometric maximum.
At bolometric maximum, the color is essentially set by the \nifs\ mass in all our models.
\label{fig_lc}
}
\end{figure*}

\begin{figure*}
\epsfig{file=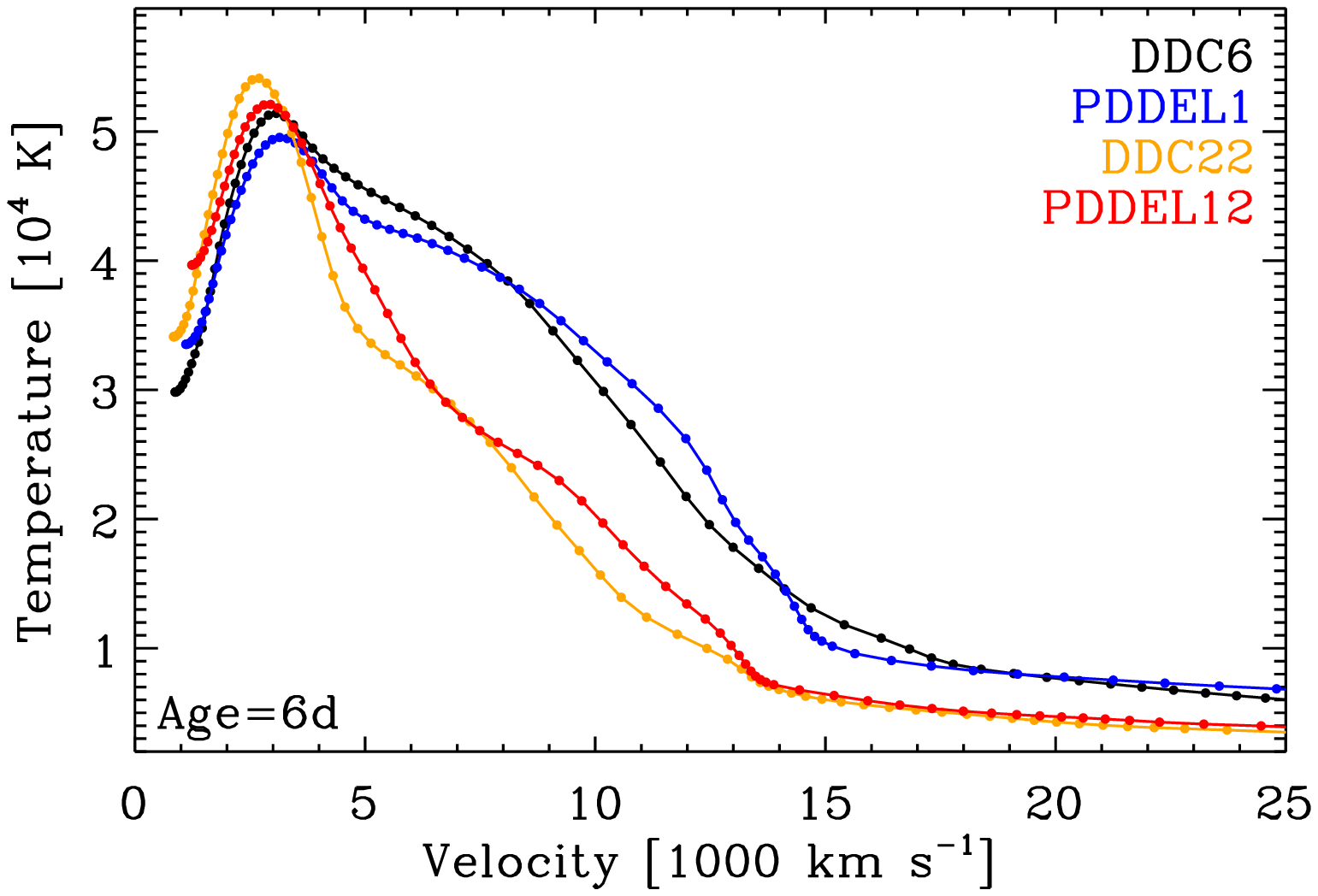,height=5.3cm}
\hspace{2mm}
\epsfig{file=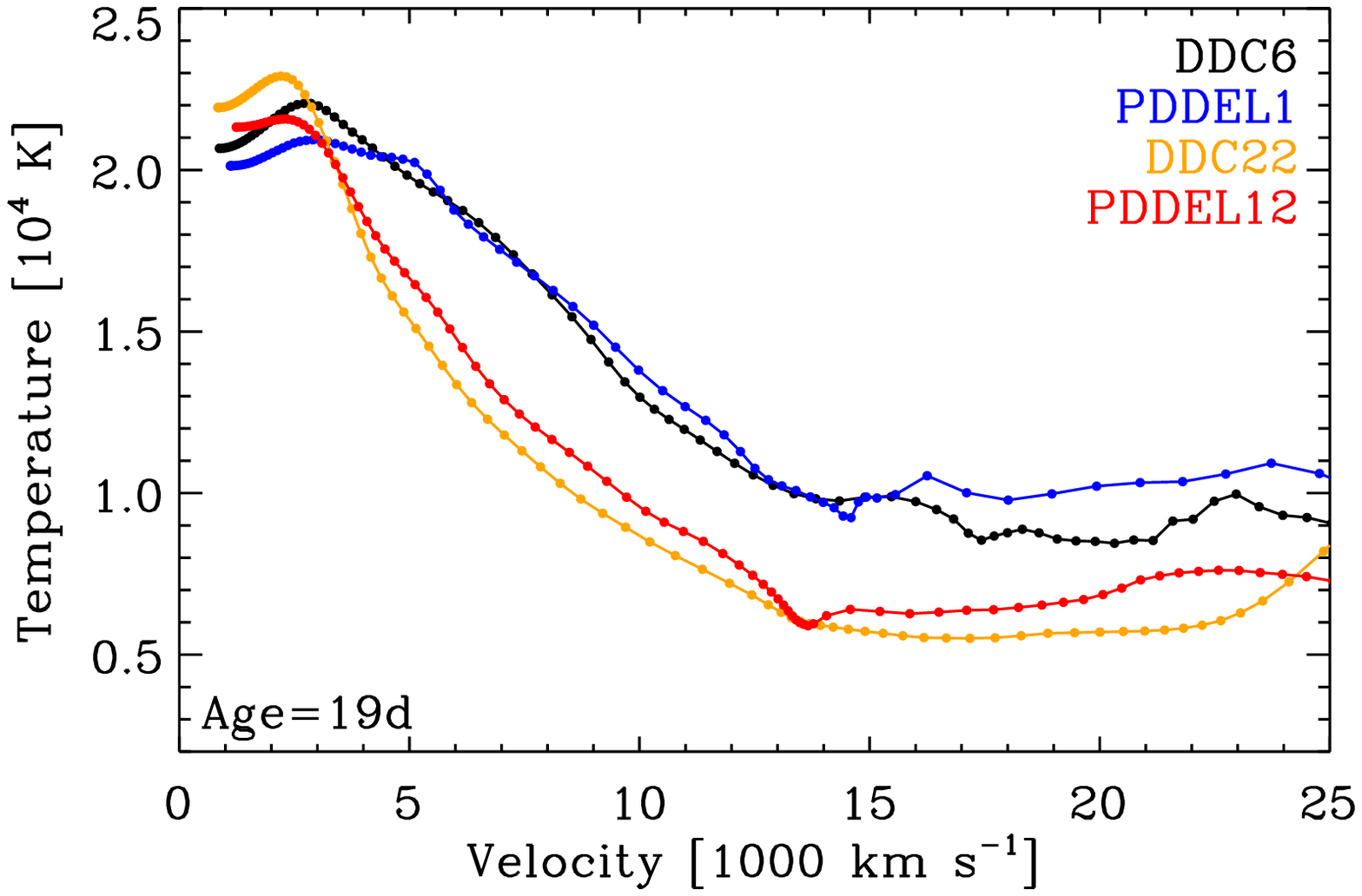,height=5.3cm}
\caption{
Comparison of the temperature structures computed by \cmfgen\ for the DDC and PDDEL
models at 6 (left) and 19\,d (right) after explosion.
At 6\,d after explosion, the location where $\tau_{\rm Rosseland}$ is 2/3
is 14000\,\kms\ in model PDDEL1, but 16700\,\kms\ in model DDC6, corresponding
to a local temperature difference of a few 1000\,\kms.
In models PDDEL12 and DDC22, that location is at 13000\,\kms\ at that time, but it
is then the sub-photospheric temperature that differs significantly.
At 19\,d after explosion, each model is around its bolometric
maximum and the temperature contrast is overall quite reduced.
At such times, it is primarily the \nifs\ mass produced in the explosion that controls
the temperature structure, rather than the initial temperature conditions.
\label{fig_comp_temp}
}
\end{figure*}

Using physically-consistent ejecta properties is essential for SN studies (see, e.g., the discussion
in \citealt{dessart_etal_13} in the context of SNe II-Plateau). Here, our simulations have to capture the
different chemical stratifications, as well as the distinct density and temperature structures of
the original DDC and PDDEL models.
As shown in the left panel of Fig.~\ref{fig_comp_tinit}, the temperature profile at 100\,s
after explosion is drastically different between models PDDEL1 and DDC6 (although they have the same
\nifs\ mass). While the temperature in \nifs-rich regions is quickly controlled by \nifs\ decay, the
temperature in \nifs-deficient regions is at early times entirely determined by the initial temperature
set by the explosion and cooling from expansion. It is much larger in the outer regions of PDDEL ejecta models
because of the ``pulsation" and associated hydrodynamic interaction.

In the right panel of Fig.~\ref{fig_comp_tinit}, we show the temperature structure at 0.5\,d
(this is the time we start the \cmfgen\ simulations), evolved
from 100\,s ignoring radiative diffusion. Although the contrast between PDDEL and DDC models
is somewhat reduced by enforcing a minimum of 6000\,K in the low-density (optically-thin)
outermost regions (see D13 for details), the temperature in the outer ejecta
is typically 10 times as large in the PDDEL models compared to the DDC ones.
The temperature shift is very large, much larger than obtained in the DDC sequence through a
change in \nifs\ mass of a factor of 5 and therefore will no doubt cause a dramatic change in
the SN radiation properties. For PDDEL models, it is essential to use this initial temperature,
which is physical, rather than assuming that it is entirely controlled by \nifs\ decay, which
is only true in \nifs-rich regions.

We show the density structure at 0.5\,d in Fig.~\ref{fig_comp_dinit}.
A large density jump is clearly visible, spanning more than an order of magnitude.
The \nifs\ is systematically confined to regions bound by this density cliff in all PDDEL models
(see Fig.~\ref{fig_comp} and Table~\ref{tab:modinfo}), and therefore plays little role in establishing
the high temperature of the outer ejecta at early times.

In contrast, the inner ejecta (say below $\sim$\,10000\,\kms) are very comparable between
the DDC and PDDEL sequences, so we anticipate that the photometric and spectroscopic differences
between the two sets of models will occur primarily at early times, during the rise to bolometric maximum.
In the next section, we discuss to what extent the different initial conditions (enhanced
temperature in the outer ejecta; presence of a density cliff at $\sim$\,13000-15000\,\kms; residual unburnt
carbon) alter the SN Ia radiative properties we obtain for the ``standard"  delayed-detonation simulations
presented in \citet{blondin_etal_13} and D13.

\section{Results}
\label{sect_res}

In this section, we discuss the photometric and spectroscopic differences between
the DDC and PDDEL series. To be concise, we focus on two sets of models at large
(DDC6 and PDDEL1) and low (DDC22 and PDDEL12) \nifs\ mass. 

\subsection{Photometric properties}

Our \cmfgen\ simulations of pulsational-delayed detonation models exhibit a number of striking
differences with their ``standard" delayed-detonation counterparts.

At times prior to peak, PDDEL models are more luminous and bluer than their corresponding DDC
model (Fig.~\ref{fig_lc}; the effect is more pronounced in $U-V$ than in the $B-R$ color shown
in the right panel). 
The early-time luminosity still correlates with \nifs\ in the PDDEL sequence,
but the sensitivity is weaker.
All these properties stem from the larger initial temperature in the outer ejecta, which is caused
not by \nifs\ but by the hydrodynamic interaction that arises in the pulsational-delayed-detonation
scenario.
In practice, assuming no decay heating, no radiative cooling, and no radiative diffusion,
the $T \propto 1/R$ evolution of outer
ejecta mass shells yields a temperature on the order of 10000\,K at 0.5\,d
for PDDEL models, and typically a factor of 10 lower in DDC models (Fig.~\ref{fig_comp_tinit}).
A temperature offset subsists for about 1-2 weeks, which suggests that the ejecta keeps a memory
of these different initial conditions (pulsation or not) until bolometric maximum
(Figs.~\ref{fig_lc}-\ref{fig_comp_temp}).

Our PDDEL models appear more luminous and bluer early on, and from
$\sim$\,1d to bolometric maximum PDDEL models span a reduced range of color compared to
our \cmfgen\ simulations of DDC models (right panel of Fig.~\ref{fig_lc}). However both
DDC and PDDEL models show similar photometric and spectroscopic properties at maximum, and post maximum.
At these times the evolution is controlled by the inner cobalt/iron rich ejecta which have
similar properties for a given \nifs\ mass. As illustrated in Figs.~\ref{fig_comp_tinit}--\ref{fig_comp_dinit}, 
the impact of the ``pulsation" is  limited to the outer ejecta layers and, consequently, to the early-time evolution.

\begin{figure}
\epsfig{file=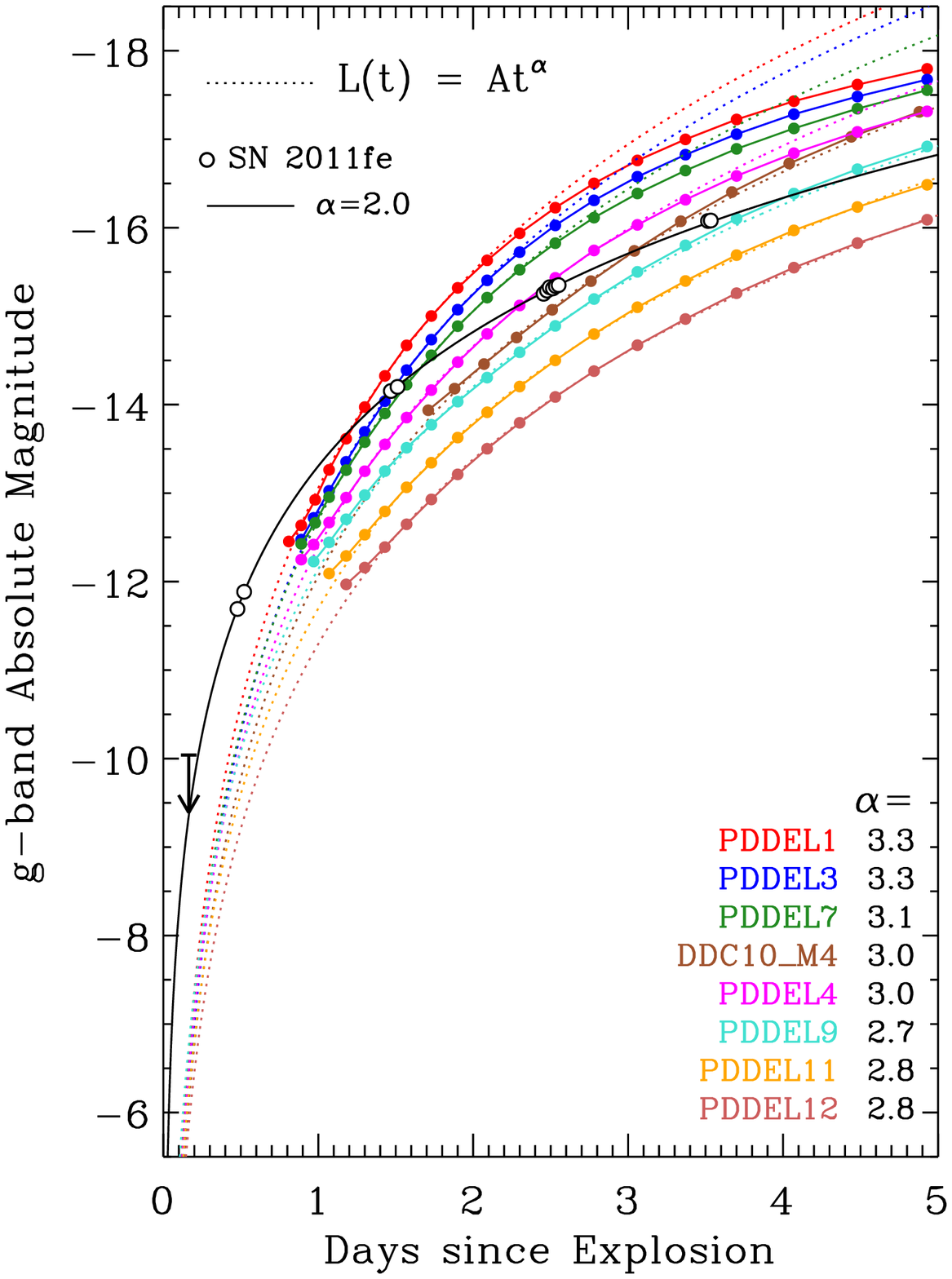,width=8.75cm}
\epsfig{file=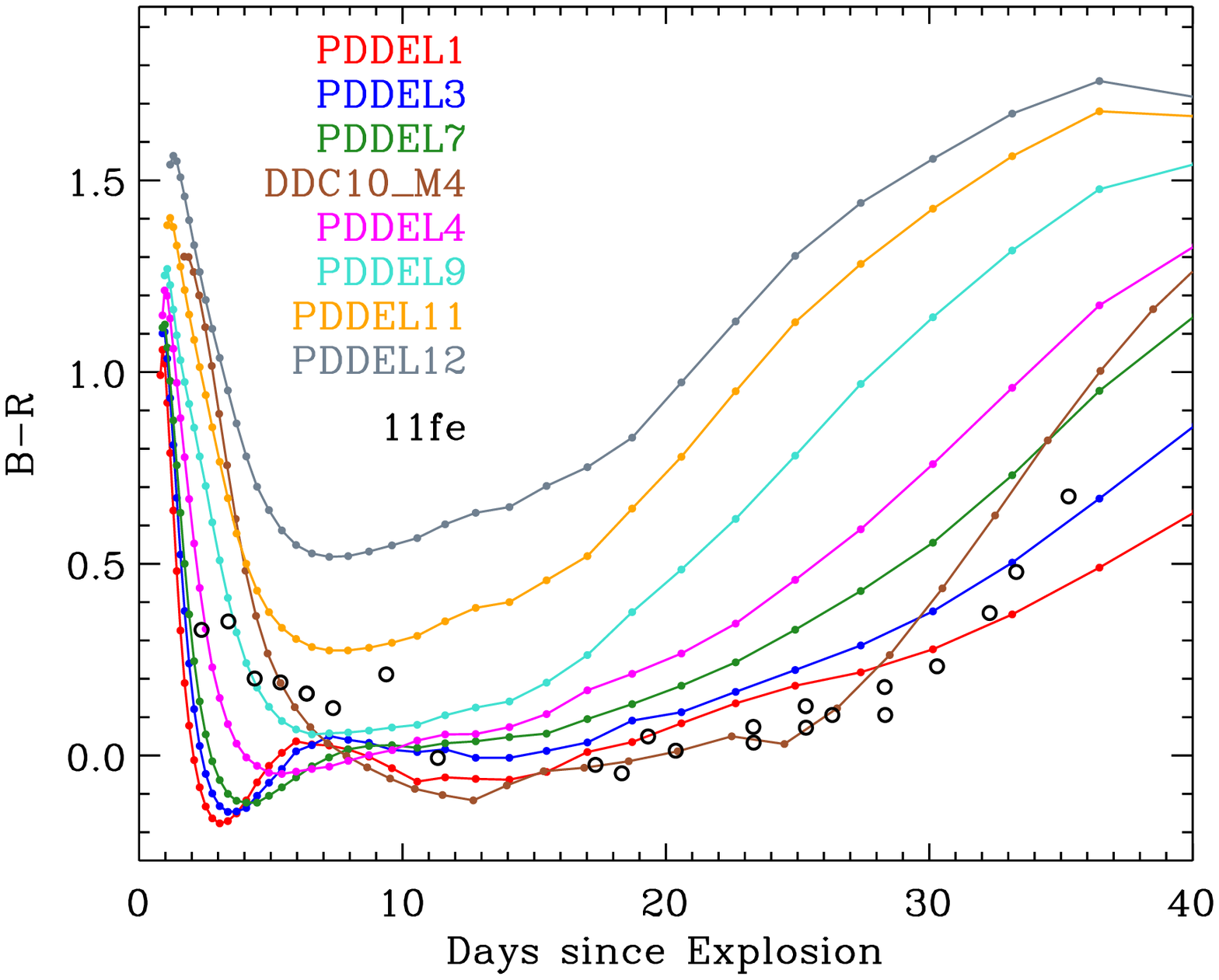,width=8.75cm}
\caption{Same as Fig.~\ref{fig_slope}, but now  for the PDDEL sequence of models.  As can be seen by a
comparison with Fig.~\ref{fig_slope}, the luminosity of the PDDEL models show a much shallower variation
of luminosity with time. In addition, the B-R colors are bluer, and show much less temporal evolution.
\label{fig_pddel_color}
}
\end{figure}

For the same level of mixing,
the early photometric behavior of PDDEL models is in better agreement with the observations of SN\,2011fe than
obtained with the DDC models (Fig.~\ref{fig_pddel_color}).
The higher luminosity and bluer color at early times makes the rising
slopes more compatible with observations, the power-law exponents now being $\sim$\,3 rather than
$\sim$\,7 for the DDC models.
PDDEL models are still too red prior to 1-2\,d, but after that those with a large \nifs\ mass match the color
of SN\,2011fe (some are even a little too blue).
Thus the pulsational-delayed-detonation mechanism offers an interesting means
to obtain bluer colors and a sizable SN Ia luminosity early after explosion, without invoking a direct contribution
from \nifs\ heating through a strong mixing of the ejecta. 
A weak correlation of the early luminosity to \nifs\ mass is present but it results not from decay heating,
but from the differing strengths of the hydrodynamic interaction that is caused by the different strengths of the detonation.

These results rely on having knowledge of the initial ejecta temperature immediately
after explosion. They depend on the use of physically consistent models rather than crafted ones with a zero initial
temperature -- it cannot be assumed that nuclear decay is solely responsible for the luminosity at early epochs.
At early times, the match would be improved by introducing a 1\,d shift, as in \citet{mazzali_etal_13},
but this worsens the match for the $B$-band rise time of model PDDEL4 with SN\,2011fe. For the purpose
of the present paper, and given the artificial setup for the ``pulsation", it is unnecessary to be more
quantitative.
The contrast between the PDDEL and DDC models suggests that a slight adjustment of the pulsation setup
and burning (through its influence on the explosion energy) could bring our PDDEL4 model in agreement
with the observations of SN\,2011fe, perhaps even using the inferred explosion time of \citet{nugent_etal_11}.
The key is that the early light curve may not be controlled primarily by nuclear decay. This works also highlights
the need for deep observations as early as possible --- such observations are needed to capture the SN when it
is about a thousand times fainter than at bolometric maximum, and to study what controls
the early evolution in luminosity and color.
  The early-time light curve and spectra provide important insights  into both the explosion 
mechanism and the progenitor properties. However, given the uncertainties in numerical models, it is important 
to build a statistical data base of SN\,Ia behavior 15--20 days before bolometric maximum  
so that we can investigate how the early-time behavior correlates with other properties of SNe Ia.

\subsection{Spectral evolution and line-profile morphology}

\begin{figure*}
\epsfig{file=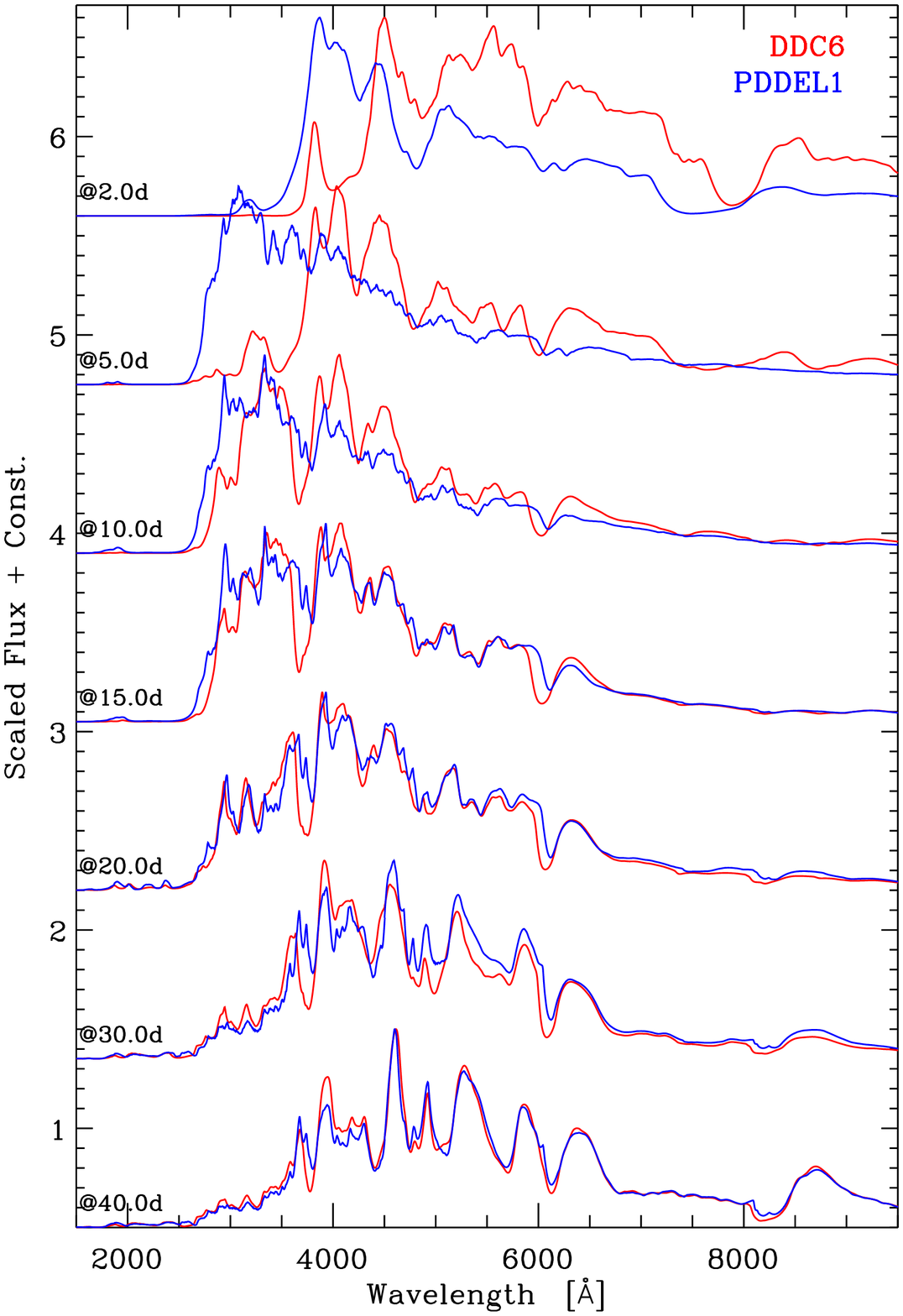,width=8.5cm}
\hspace{2mm}
\epsfig{file=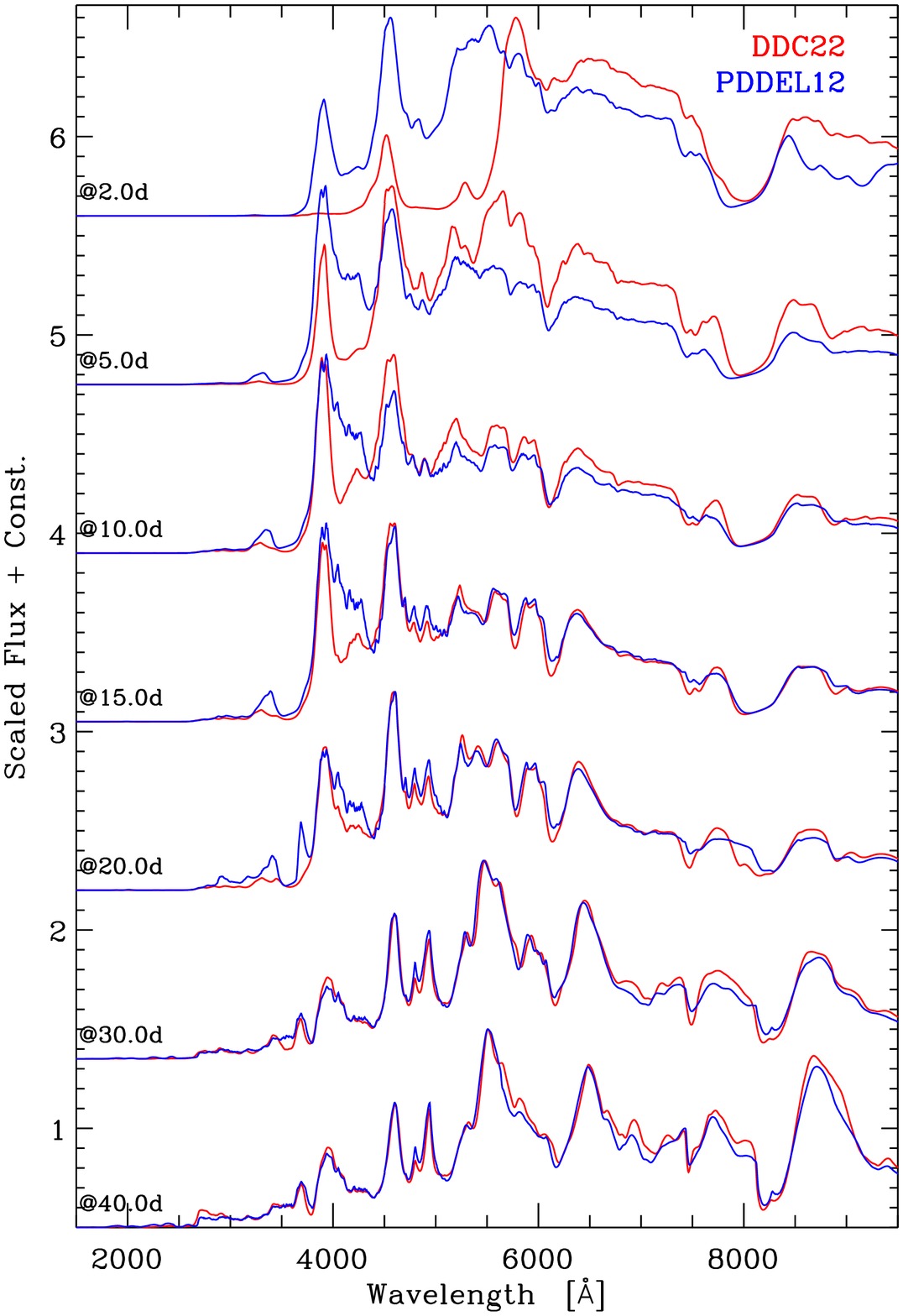,width=8.5cm}
\caption{
Spectral comparison for each pair of DDC/PDDEL models having comparable
\nifs\ mass of $\sim$\,0.75\,\msun\ (DDC6 and PDDEL1) and $\sim$\,0.25\,\msun\ (DDC22 and PDDEL12).
Time increases downward, starting at 2\,d after explosion.
While the spectral contrast at early time is huge,
each model matches closely its counterpart at and beyond bolometric maximum, which occurs
at $\sim$\,20\,d after the explosion).
\label{fig_spec_comp}}
\end{figure*}

Spectroscopic properties of PDDEL models are markedly different from
those of DDC models up until (approximately) bolometric maximum.
A montage of spectra that illustrates these spectroscopic differences for our two sets of
PDDEL/DDC models at large and small \nifs\ mass is shown in Fig.~\ref{fig_spec_comp}.
At early times, the bluer color in PDDEL models is associated with a bluer spectrum
throughout the optical range, although all models are typically too cold to radiate
much flux in the UV (i.e., shortward of $\sim$\,3000\,K).
The concept of a photosphere is still meaningful at such times so this change in color
and slope stems from the higher temperature of the radiating layer, in analogy to
the emission from a blackbody.

\begin{figure*}
\epsfig{file=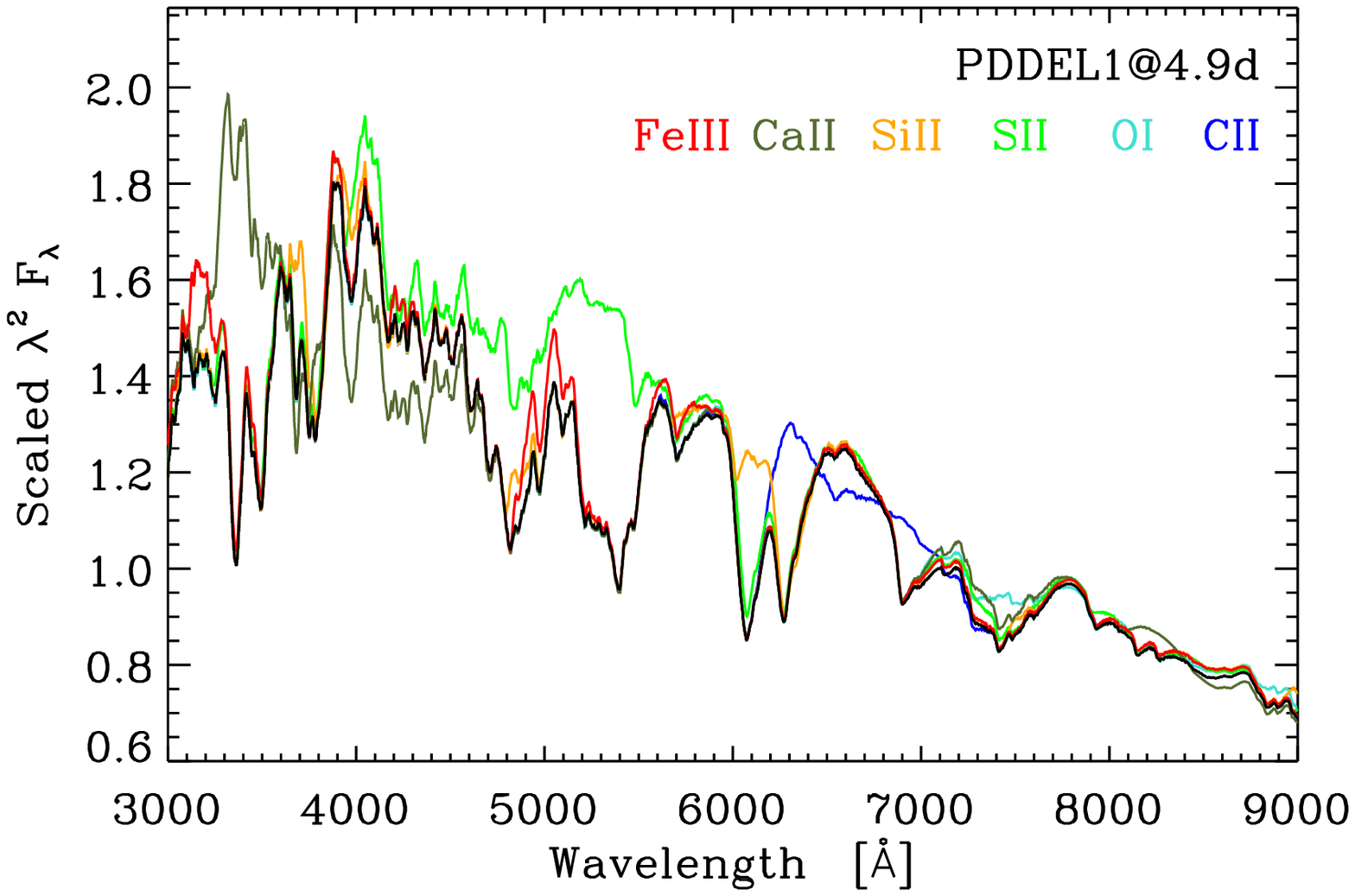,width=8.5cm}
\epsfig{file=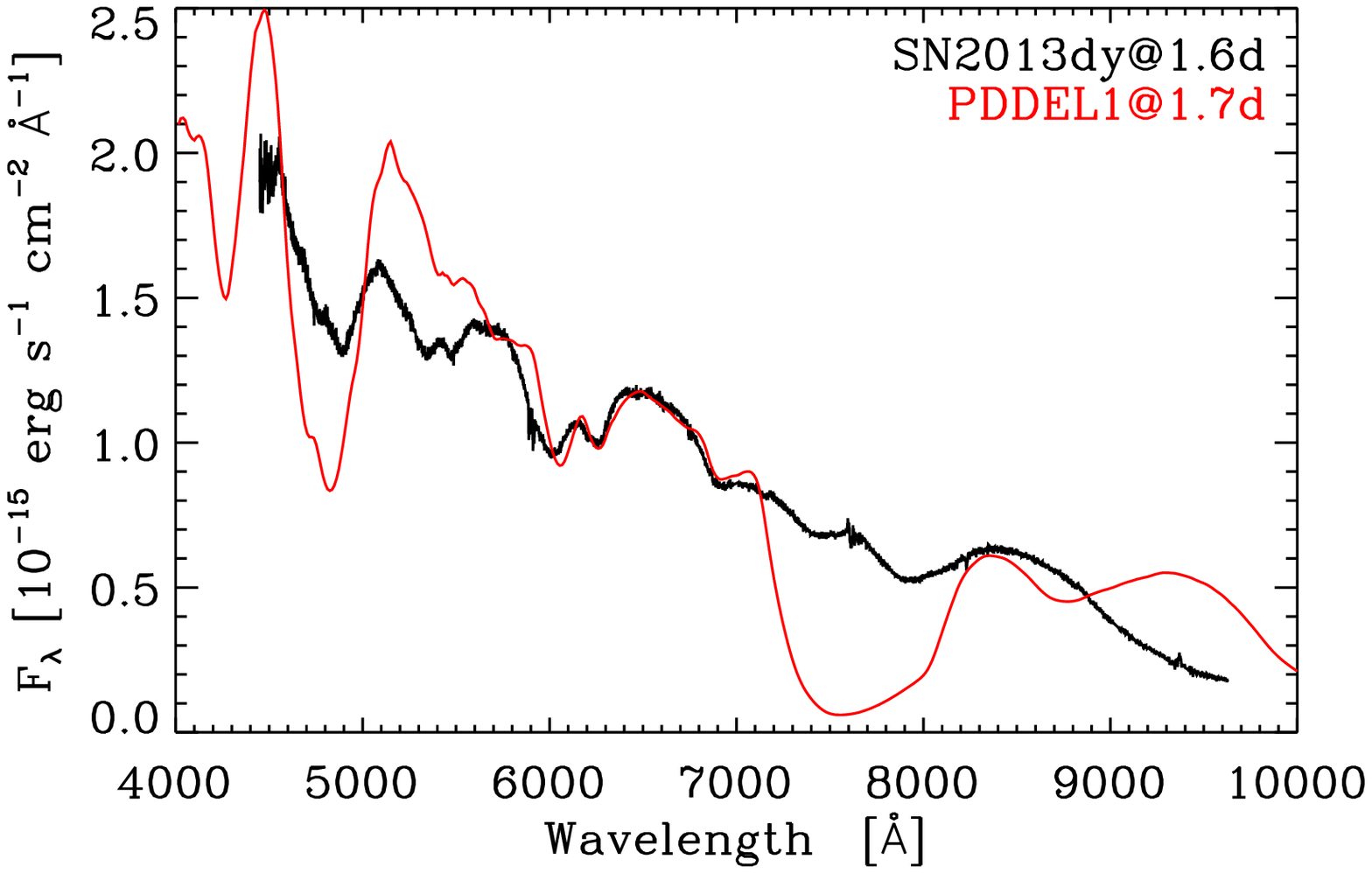,width=8.5cm}
\caption{{\bf Left:} Synthetic spectrum (black) for model PDDEL1 at 4.9\,d after explosion,
together with the model predictions where selected species are excluded from
the calculation (other colors).  For better visibility, we show the quantity $\lambda^2 F_{\lambda}$.
As expected, such pulsational-delayed-detonation synthetic spectra are
dominated by lines from IMEs (in particular, Ca\two, S\two, and Si\two), rather than IGEs. 
Moderate line blanketing from Fe\three\ is visible in this very blue spectrum, as well as two distinct features
associated with C\two.
{\bf Right:} Comparison between the observations of SN\,2013dy at 1.6\,d \citep{zheng_etal_13} and the
model PDDEL1 at 1.7\,d. The model, which is corrected for reddening and distance dilution
using the parameters of Zheng et al., is scaled by a  factor 1.08 to match the observed flux near 6000\,\AA.
Despite clear problems with the width of strong lines (e.g., Ca\two\ triplet at 8500\,\AA;
we however find that the strong line features quickly weaken and become narrower, see left panel),
the C\two\ and Si\two\ lines, which form in the vicinity of the photosphere, are well reproduced.
\label{fig_pddel_spec@5d}}
\end{figure*}

The higher temperature in the outer ejecta of PDDEL models leads to an increase
in ionization at early times. This, together with the presence of unburnt carbon, produces
clearly visible C\two\ lines at early times in PDDEL models,
in particular for high \nifs\ mass (Fig.~\ref{fig_pddel_spec@5d}).
Besides the doublet at 6580\,\AA, the C\two\,7234\,\AA\ triplet is also present.
This synthetic spectrum exhibits a striking similarity with the observations of
SN\,2013dy \citep{zheng_etal_13}, although our model is not particularly abundant in carbon
(cumulative mass of 0.02\,\msun). Instead, their visibility is caused by the enhanced ionization and
the presence of carbon at relatively small velocities, just above the density cliff where the spectrum forms
at early times.
Blanketing by Fe\three\ lines is visible but quite weak. The conspicuous presence of lines from
Si\two, S\two, and other IMEs indicates that such PDDEL models do not match
the basic properties of 91T-like events (see, e.g., \citealt{filippenko_etal_92}).

It is clear from the early time spectrum that the ejecta model properties are in error at 
large velocities. Strong lines like the Ca\two\,8500\,\AA\ triplet are very broad in the model but
much narrower in the observations. Although a mismatch in ionization is a possibility, an alternative
is that the density is much lower in reality beyond $\sim$25000\,\kms\ than our hydrodynamical code
presently predicts. This applies to both DDC and PDDEL models. 
\footnote{We make the reasonable assumption that the abundances in the outer layers are solar.  
The validity of this assumption will depend on the SN model and the explosion model. 
We note, for example, that an isolated white dwarf has very low metal abundances at the surface due
to gravitational settling of heavy elements in the atmosphere.}

 The density cliff,  present in PDDEL ejecta but absent in  ``standard" delayed-detonation models,
is central to the understanding of the early spectral morphology of PDDEL models.
Because the outer ejecta layers have a very low density above the cliff, the spectrum formation
region quickly recedes after explosion to $\sim$\,14000\,\kms\ where the density steeply rises
in our PDDEL set (Fig.~\ref{fig_comp_dinit}).
While the spectrum formation region tends to continuously recede in DDC models, it resides
for longer within that cliff because it represents a large jump in optical depth.
This density jump also spans a very narrow range of velocities, which drastically reduces
the change of line widths with time.
Consequently, the radiation forms and emerges from a region with a large density gradient,
which is known to reduce the spatial extent of line formation,
and thus produce weak absorption/emission in SN ejecta \citep{DH05a,DH05b}.
Because of the particular chemical stratification of PDDEL models in velocity space,
line absorption (especially associated with IMEs) is confined to a narrow velocity range (Fig.~\ref{fig_comp}),
causing narrow absorption troughs.
Interestingly, the difference in the profile morphology of Si\two\,6355\,\AA\ is strong
between models DDC6 and PDDEL1, even though both ejecta have the same total mass of Si.
{\it This shows that different line strengths can arise without associated abundance changes.}

Overall, pulsational-delayed-detonation models tend to produce pre-peak spectra with weak lines
whose absorption maxima trace a modest range of velocities. This contrasts with ``standard"
delayed-detonation models which exhibit stronger lines with larger velocity changes in P-Cygni
profile minima with time (see Section~\ref{sect_vg}).

\subsection{A pulsational-delayed-detonation model for SN\,2011\lowercase{fe}}

   In this section, we make a more specific comparison between our PDDEL models
and SN\,2011fe. In agreement with the earlier studies of \citet{pereira_etal_13} and
\citet{mazzali_etal_13}, we find that an ejecta with 0.53\,\msun\ of \nifs\ satisfactorily
matches the peak luminosity, 

  We show a comparison between the SN\,2011fe $B$-band light curve and
model PDDEL4 in Fig.~\ref{fig_mag_11fe}. The model matches the peak magnitude
and the decline rate for the first 20\,d after maximum (beyond that, SN\,2011fe fades faster than
the model). A slight offset remains at early times, but as we argue above,
modest adjustments in the setup for the ``pulsation'' suggest the two could probably be brought
into agreement. In this figure, we add the observations of SN\,2002bo and model
DDC15 (a detailed study of SN\,2002bo, with a comparison with model DDC15, will be
presented in Blondin et al., in prep.).
The light curves for the two models and the two SNe lie very close to each other, although SN\,2002bo
is notorious for its broad lines
\citep{benetti_etal_04}, while SN\,2011fe has in contrast narrow lines. This demonstrates
that for a given light curve, constrained primarily by the \nifs\ mass, spectral diversity
can arise  even from spherical ejecta like the PDDEL and DDC models we study here.
Interestingly, the strongly mixed model DDC10\_M4 and the pulsation-delayed-detonation
model PDDEL4 have comparable early-time light curves.

As mentioned earlier, the match at early times would be improved by introducing
a 1\,d shift, as in \citet{mazzali_etal_13}, but this worsens somewhat the match for the $B$-band rise time
of model PDDEL4. More importantly, it seems that the need to argue for a dark phase in a
SN Ia stems from the assumption that the SN radiation is powered entirely by \nifs.
At early times, this radiation could instead come from shock-deposited energy in a
configuration conceptually equivalent to the pulsational-delayed-detonation presented here.
In that sense, SNe Ia could exhibit a short but not so faint post-breakout phase, analogous
to what is expected in SNe IIb/Ib/Ic \citep{dessart_etal_11}.
This highlights the need for more sensitive surveys, to push the detection limits.

   Spectroscopically, the match between model PDDEL4 and SN\,2011fe is also good
   (Fig.~\ref{fig_spec_11fe}),
   especially if we consider that there is no adjustment made to the initial explosion
   model (which we take at $\sim$\,100\,s after explosion). In particular, unlike \citet{mazzali_etal_13},
   we do not adjust any of the species mass fractions or the density profile. Despite the lack of
   freedom, the basic color evolution is matched, and line widths are also well matched (we synchronize
   the model and the observations at $B$-band maximum in this montage). We also predict the presence of
   C\two\,6580\,\AA\ and 7234\,\AA\ for the first few days after explosion.
   While these are stronger in the model than observed in SN\,2011fe, they are
   in close agreement with the C\two\ lines in SN\,2013dy.

   The largest discrepancies are at early times, but they seem to concern primarily the
   widths of the lines. It is hard at this stage to address this since the outer ejecta
   is poorly covered by the hydrodynamical code, which probably compromises the
   accuracy of the density profile beyond 20000-30000\,\kms. This density profile
   would also be different if we changed the setup for the pulsation. A steeper profile would
   quench absorption/emission at large velocity and would prevent the formation
   of the very broad Ca\two\ features that we predict, for example. Because of this and uncertainties on
   the temperature/ionization, abundance determinations for these layers are uncertain.
   Our model has an equal share of C and O beyond 15000\,\kms\ and matches reasonably well the
   C\two\ and O\one\ lines. With such abundances, the strength of these lines in our simulations is primarily
   affected by ionization. We do not find strong evidence for overabundance in carbon, as proposed by
   \citet{mazzali_etal_13} for SN\,2011fe.

   In Fig.~\ref{fig_spec_11fe}, we also include a comparison of model PDDEL4 with the observations
   of SN\,2011fe at +82\,d after $B$-band maximum (unfortunately, the data from \citealt{pereira_etal_13}
   has a 50\,d gap and there is at present no public data available during that gap).
   The spectral morphology is well reproduced, in particular the relative flux distributions, line strengths
   and widths, with only a slight offset in absolute flux (0.2\,mag in $B$). This suggests that apart from
   a few mismatches at early times (Ca\two\,H\&K region, red part of the optical), the pulsational-delayed-detonation
   model PDDEL4 is a sound physical model of SN\,2011fe. Of course, this does not mean it is
   the only possible ejecta  that has a similar evolution as SN\,2011fe, but it strongly suggests
   that delayed detonations of that nature are not incompatible with this event.
   This result contrasts with the conclusions of \citet{roepke_etal_12}, who favor a white-dwarf---white-dwarf merger
   progenitor based on maximum-light spectral properties.
   However, the numerical procedure we employ to set the pulsation may actually mimic with
   some basic fidelity the effect that a buffer of mass, resulting from a merger event, would have on the 
   exploding white dwarf remnant. 

\subsection{HVG versus LVG SNe I\lowercase{a}}
\label{sect_vg}

The hydrodynamic interaction that takes place in pulsational-delayed-detonation models
of SNe Ia causes two features that affect the evolution of line profile widths.
First, the spectrum formation region will tend to reside longer within the density cliff located
at the outer edge of the IME-rich layers. Secondly, the bulk of the ejecta mass covers a reduced
range of velocities, in particular for IMEs.
Together, these effects reduce the maximum range of line-profile widths as the SN ages.
These effects are absent in ``standard" delayed-detonation models.\footnote{The hydrodynamic
interaction taking place in pulsational-delayed-detonation models is strong and would likely
be present in a multi-D simulation started with the same conditions. This, however, remains
to be studied.}

\begin{figure*}
\epsfig{file=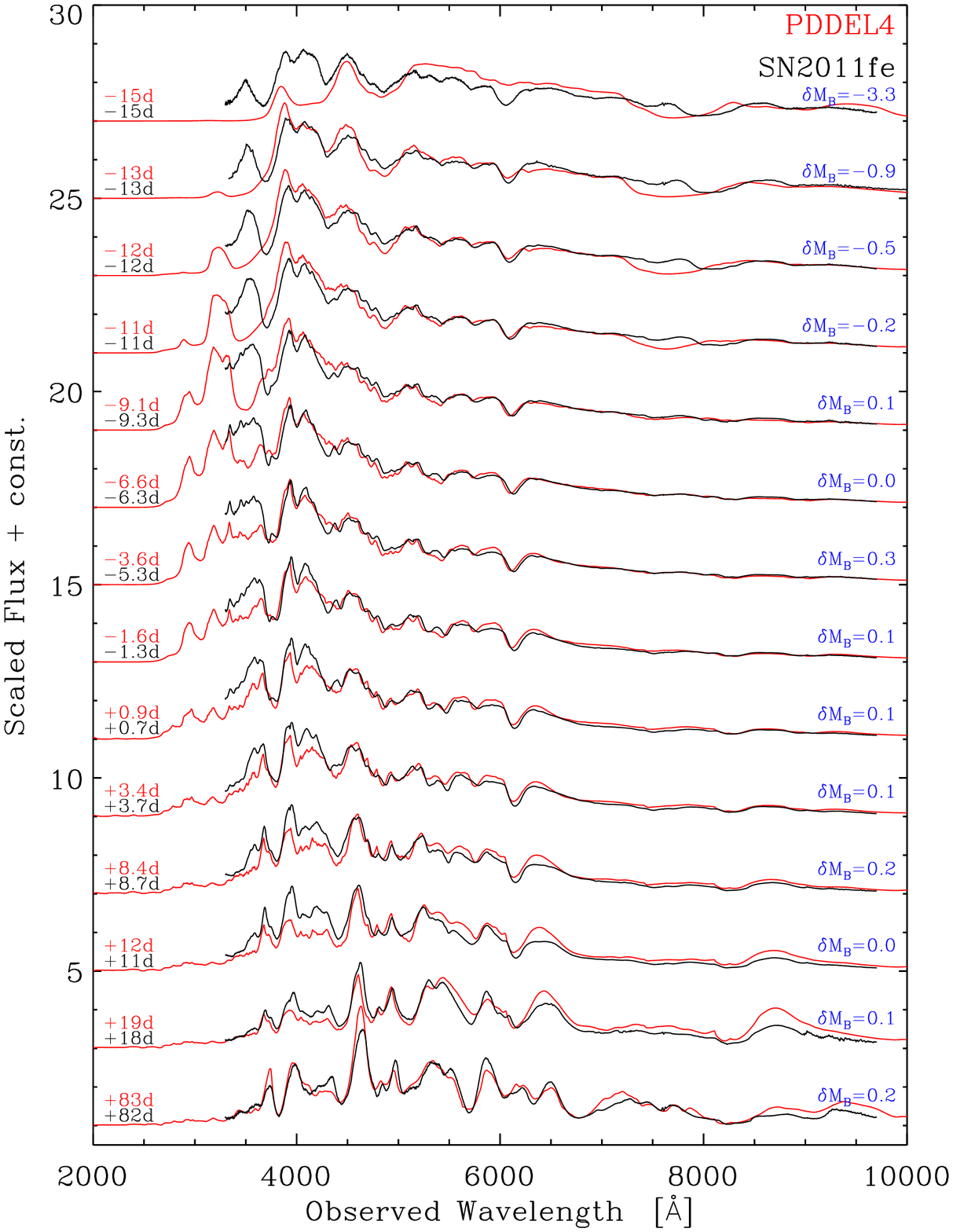,width=17cm}
\caption{
Comparison between model PDDEL4 and the observed spectra of SN\,2011fe.
Times are given with respect to $B$-band maximum.
We correct the synthetic flux to account for the distance, redshift, and extinction of SN\,2011fe.
Spectra are scaled vertically to facilitate spectral comparisons, although the label
on the right gives the true $B$-band magnitude offset between model and
observations at each date.
\label{fig_spec_11fe}}
\end{figure*}

\begin{figure}
\epsfig{file=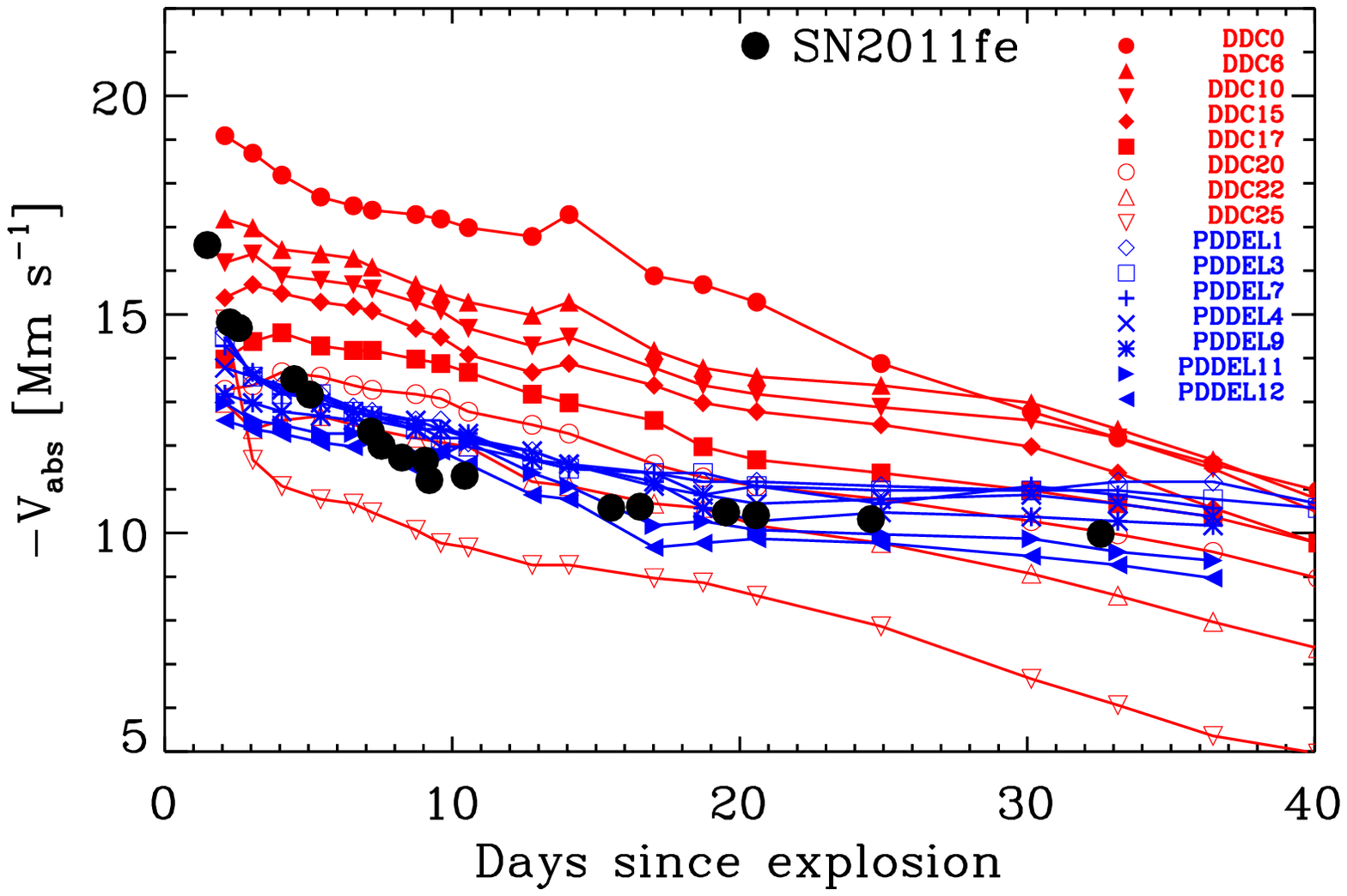,width=8.75cm}
\caption{Illustration of the evolution of the Doppler velocity at maximum
absorption in the P-Cygni profile associated with the Si\two\,6355\,\AA\ doublet
line for the PDDEL models (blue) and DDC models (red).
Extra symbols differentiate each model within each set. The PDDEL models
show a much smaller variation with time than do the DDC models.
Black dots correspond to the corresponding measurement in SN\,2011fe.
\label{fig_line_path}}
\end{figure}

Hence, the distinct explosion mechanism between PDDEL and DDC models
offers one physical source of spectral diversity. As shown in Fig.~\ref{fig_line_path},
these two populations of models are reminiscent of the so-called HVG and LVG SNe Ia
\citep{benetti_etal_05}.
While DDC models can match the broad Si\two\,6355\,\AA\ trough in
SNe Ia like 2009ig or 2002bo \citep{blondin_etal_13}, they poorly  fit  the
SNe Ia exhibiting a narrower Si\two\,6355\,\AA\ line, such as SN\,2005cf (D13), or SN\,2011fe.
For these, the PDDEL models match both the narrow absorption and
the near-constancy of the location of the absorption maximum (Fig.~\ref{fig_spec_11fe}).
More quantitatively, over the range 20 to 30\,d after explosion, we obtain a mean velocity
gradient of 100--300\,km\,s$^{-1}$\,d$^{-1}$ (absolute value)
for models DDC and $\sim$\,0\,km\,s$^{-1}$\,d$^{-1}$ for models PDDEL.

In the context of delayed-detonation and pulsational-delayed-detonation models,
the HVG and LVG classes of SNe Ia no longer require strong ejecta asphericity,
as proposed by  \citet{maeda_etal_10} or discussed by \citet{blondin_etal_11}.
They may instead arise from distinct but quasi-spherical mass distributions in velocity space,
some SNe Ia having a larger mass at large velocity (e.g., SN\,2009ig) than others
(e.g., SNe 2011fe or 2005cf).

\section{Conclusions}
\label{sect_conc}

In this work, we have investigated the early-time behavior of delayed-detonation models. Although we
obtain good agreement around bolometric maximum for these models, we find
that the models tend to be very faint and red during the first week after explosion
compared to the observations of SN\,2011fe.

Assuming that decay heating is the key mechanism for powering SN Ia radiation,
we investigate the effect of treating decay chains other than that of \nifs.
While \nifs\ is primarily produced at high density, a number of unstable isotopes, 
especially associated with IMEs, are produced in the outer ejecta. We find that 
these additional decays enhance the early-time luminosity, 
but at a level that is much too small to reconcile our DDC model with observations of SN 2011fe.

Secondly, we examined the role of chemical mixing in our reference model DDC10 (D13).
As expected, we find that radial microscopic mixing, in particular of \nifs, influences the early time properties
of these models, leading to a luminosity increase of a factor of 2-3 for the model
with the strongest mixing. This also produces bluer colors and partially reconciles the model photometry
with the observations of SN\,2011fe.
However, mixing tends to enhance line-optical depths at early times due to larger
metal mass fractions and excess heating at large velocities. This enhances line broadening
and exacerbates the discrepancy with SN\,2011fe, which shows relatively narrow and weak lines.

As an alternative to the DDC models, we have presented results from explosion models  named
pulsational-delayed detonations. Although our setup for the explosion is artificial,
it captures approximately what may occur if SNe Ia arise from the explosion of a
white dwarf in a pulsation cycle, or from the explosion following the merger of two white dwarfs, i.e.,
a white dwarf remnant surrounded by a buffer of mass.
In this context, pulsational-delayed detonations differ from standard delayed detonation
models by the different chemical stratification of their ejecta in velocity space,
the greater survival of unburnt carbon ($\sim$\,0.02\,\msun\ in all our PDDEL models, irrespective
of \nifs\ mass), the presence of a density cliff at the outer edge of the
IME-rich layers, and the much higher initial temperature above (at higher velocities than) the density cliff.

Consequently, the radiation properties of pulsational-delayed detonations have
several striking features absent in standard delayed-detonation models in which the same chemical 
mixing is applied.
The early time model luminosity can be increased by up to a factor of 10, and the color
is significantly bluer at early times, in much closer agreement with SN\,2011fe than obtained with weakly mixed
DDC models.
These differences do not stem from the influence of \nifs, but instead from the different  temperature
of the outer-ejecta that was shock-heated to very high temperatures in the initial
phases of the explosion.
Other interesting features of the pulsational-delayed detonations, which
are seen in numerous SNe Ia, are
\vspace{-5pt}
\begin{enumerate} 
\item
the weakness of line
profiles at early times, making the spectra more featureless;
\item
the presence of C\two\ lines
for up to a week, with stronger features for SNe Ia having a larger \nifs\ mass
(an ionization rather than an abundance effect);
\item
 the markedly different morphology of line profiles. The Si\two\,6355\,\AA\ profile
 exhibits narrower P~Cygni absorption whose velocity at maximum absorption varies
 only weakly (as compared to the DDC models) over time.
\end{enumerate}
\vspace{-5pt}

While some of these features are reminiscent of 91T-like objects (see, e.g., \citealt{filippenko_etal_92}),
the presence at early times of lines from C/O and IMEs suggests pulsational-delayed detonations
are not the explanation for these 91T-like events. However, we find that our
pulsational-delayed detonation model PDDEL4 matches satisfactorily the multi-epoch spectra and
muti-band light curves of SN\,2011fe, including line features, line profile morphology, trajectory
in velocity space of absorption minima. Our PDDEL models are close to matching
the $t^2$ increase in brightness of SN\,2011fe, as well as the relatively modest evolution in color
on the way to bolometric maximum. Although standard delayed detonation models with strong mixing 
also provide a fair agreement to the early time photometry,  they produce broader lines
than observed in SN\,2011fe.

Such pulsational-delayed detonation models break a number of intuitive but potentially
wrong notions. First, the SN Ia radiation does not necessarily stem exclusively from decay
heating, compromising ``diffusion'' models that attempt to explain the origin of the early light curve
of SNe Ia.
Second, the variations in line-profile width are not exclusively sensitive to variations in explosion
energy, since some kinetic energy may be channeled into internal/radiation energy.
If so, and in the present context, broad and narrow line SNe Ia should have different
colors at early time. In this context we highlight SN\,2011fe and SN\,2002bo,
which have similar light curves but very different spectral line profiles.
More generally, pulsational-delayed detonations and delayed detonations produce
two distinct SN Ia populations reminiscent of the HVG and LVG SNe Ia \citep{benetti_etal_05}.

Interestingly, this spectral diversity is not the result of ejecta asphericity \citep{maeda_etal_10}.
It may be that in some circumstances, these asphericities are modest and it is instead
the global, quasi-spherical, ejecta mass distribution that is fundamentally altered in certain
explosion configurations. Thus spectral  diversity may not  result from a randomization
of viewing angles on a 3-D structure but from the random sampling of intrinsically diverse
quasi-spherical SN Ia ejecta. Truth is likely to lie in between these two standpoints,
with both multi-D effects present together with SNe Ia having fundamentally distinct
angle-averaged structures.

C\two\ lines are present in our early-time spectra of PDDEL models (yielding an interesting
agreement with the observations of SN\,2013dy; \citealt{zheng_etal_13}), primarily because of the
enhanced ejecta temperature from the hydrodynamic interaction, but they are systematically absent
in the corresponding spectra of DDC models. This suggests a natural association of C\two\ detection
and LVG SNe Ia. This correlation has been observed by \citet{parrent_etal_12}.

A weak interaction/pulsation or the presence of a low mass buffer around the exploding white dwarf can
dramatically alter the outer ejecta layers of the SN Ia explosion. 
However, such modulations do not impact the bulk
of the ejecta mass (for a given \nifs\ mass, our DDC/PDDEL models have  essentially the same structure
inside $\sim$\,12000\,\kms), and so this ``pulsation" configuration can bring diversity to
the early-time evolution of SNe Ia without impacting much the maximum and
post-maximum evolution (Fig.~\ref{fig_lc}), which is primarily influenced in our set of DDC and
PDDEL models by the total \nifs\ mass.
A specific discussion on the decline rate of all our DDC and PDDEL models, in connection with the
width-luminosity relation, is left to a future study.

\section*{Acknowledgments}

LD and SB acknowledge financial support from the European Community through an
International Re-integration Grant, under grant number PIRG04-GA-2008-239184,
and from ``Agence Nationale de la Recherche" grant ANR-2011-Blanc-SIMI-5-6-007-01.
DJH acknowledges support from STScI theory grant HST-AR-12640.01, and NASA theory grant NNX10AC80G.
This work was also supported in part by the National Science Foundation under Grant No. PHYS-1066293 and
benefited from the hospitality of the Aspen Center for Physics.
AK acknowledges the NSF support through the NSF grants AST-0709181 and TG-AST090074.
This work was granted access to the HPC resources of CINES (France) under the
allocation c2013046608 made by GENCI (Grand Equipement National de Calcul Intensif).

\appendix

\setcounter{table}{2}

\begin{table}
\begin{center}
\caption{
Summary of the 2-step decay chains used.
For all chains, we give the characteristics for each
of the two steps, starting with the half-life, the total
energy radiated in $\gamma$-rays $Q_{\nu}$, and  the total
energy liberated in the form of particles $Q_{\rm th}$.
We then list the main $\gamma$-ray lines emitted in each decay
together with their probability.
\label{tab_nuc1}}
\vspace{0.3cm}
\begin{tabular}{cccc}
\hline
 \multicolumn{4}{c}{  $\iso{56}Ni \rightarrow \iso{56}Co \rightarrow \iso{56}Fe$ } \\
   \multicolumn{2}{c}{$\iso{56}Ni \rightarrow \iso{56}Co$  }  &   \multicolumn{2}{c}{$\iso{56}Co \rightarrow \iso{56}Fe$  }  \\
\hline
  \multicolumn{2}{c}{$t_{1/2}=$          6.075      d} &  \multicolumn{2}{c}{$t_{1/2}=$         77.233     d} \\
        \multicolumn{2}{c}{$Q_{\gamma}=$          1.718    MeV} &        \multicolumn{2}{c}{$Q_{\gamma}=$          3.633   MeV} \\
        \multicolumn{2}{c}{$Q_{\rm th}=$          0.000    MeV} &        \multicolumn{2}{c}{$Q_{\rm th}=$          0.116   MeV} \\
\hline
       $E_\gamma$ &             Prob. &        $E_\gamma$ &             Prob. \\
     0.158  &      98.8  &     0.511  &      38.0   \\
     0.270  &      36.5  &     0.847  &     100.0   \\
     0.480  &      36.5  &     0.977  &       1.4   \\
     0.750  &      49.5  &     1.038  &      14.0   \\
     0.812  &      86.0  &     1.175  &       2.3   \\
     1.562  &      14.0  &     1.238  &      67.6   \\
            &            &     1.360  &       4.3   \\
            &            &     1.771  &      15.7   \\
            &            &     2.015  &       3.1   \\
            &            &     2.035  &       7.9   \\
            &            &     2.598  &      17.3   \\
            &            &     3.010  &       1.0   \\
            &            &     3.202  &       3.2   \\
            &            &     3.253  &       7.9   \\
            &            &     3.273  &       1.9   \\
\hline
 \multicolumn{4}{c}{  $\iso{57}Ni \rightarrow \iso{57}Co \rightarrow \iso{57}Fe$ } \\
     \multicolumn{2}{c}{$\iso{57}Ni \rightarrow \iso{57}Co$}  &     \multicolumn{2}{c}{$\iso{57}Co \rightarrow \iso{57}Fe$}  \\
\hline
  \multicolumn{2}{c}{$t_{1/2}=$          1.483      d} &  \multicolumn{2}{c}{$t_{1/2}=$        271.740     d} \\
        \multicolumn{2}{c}{$Q_{\gamma}=$          1.937    MeV} &        \multicolumn{2}{c}{$Q_{\gamma}=$          0.122   MeV} \\
        \multicolumn{2}{c}{$Q_{\rm th}=$          0.154    MeV} &        \multicolumn{2}{c}{$Q_{\rm th}=$          0.000   MeV} \\
\hline
       $E_\gamma$ &             Prob. &        $E_\gamma$ &             Prob. \\
     0.127  &      16.7  &     0.014  &       9.2   \\
     0.511  &      87.0  &     0.122  &      85.6   \\
     1.378  &      81.7  &     0.137  &      10.7   \\
     1.758  &       5.8  &            &             \\
     1.919  &      12.3  &            &             \\
\hline
 \multicolumn{4}{c}{   $\iso{48}Cr \rightarrow \iso{48}V \rightarrow \iso{48}Ti$ } \\
      \multicolumn{2}{c}{$\iso{48}Cr \rightarrow \iso{48}V$}  &      \multicolumn{2}{c}{$\iso{48}V \rightarrow \iso{48}Ti$}  \\
\hline
  \multicolumn{2}{c}{$t_{1/2}=$          0.898      d} &  \multicolumn{2}{c}{$t_{1/2}=$         15.973     d} \\
        \multicolumn{2}{c}{$Q_{\gamma}=$          0.432    MeV} &        \multicolumn{2}{c}{$Q_{\gamma}=$          2.910   MeV} \\
        \multicolumn{2}{c}{$Q_{\rm th}=$          0.002    MeV} &        \multicolumn{2}{c}{$Q_{\rm th}=$          0.145   MeV} \\
\hline
       $E_\gamma$ &             Prob. &        $E_\gamma$ &             Prob. \\
     0.112  &      96.0  &     0.511  &      99.8   \\
     0.308  &     100.0  &     0.944  &       7.8   \\
     0.511  &       3.2  &     0.984  &     100.0   \\
            &            &     1.312  &      97.5   \\
            &            &     2.240  &       2.4   \\
\hline
 \multicolumn{4}{c}{   $\iso{49}Cr \rightarrow \iso{49}V \rightarrow \iso{49}Ti$ } \\
      \multicolumn{2}{c}{$\iso{49}Cr \rightarrow \iso{49}V$}  &      \multicolumn{2}{c}{$\iso{49}V \rightarrow \iso{49}Ti$}  \\
\hline
  \multicolumn{2}{c}{$t_{1/2}=$          0.029      d} &  \multicolumn{2}{c}{$t_{1/2}=$        330.000     d} \\
        \multicolumn{2}{c}{$Q_{\gamma}=$          1.055    MeV} &        \multicolumn{2}{c}{$Q_{\gamma}=$          0.000   MeV} \\
        \multicolumn{2}{c}{$Q_{\rm th}=$          0.598    MeV} &        \multicolumn{2}{c}{$Q_{\rm th}=$          0.000   MeV} \\
\hline
       $E_\gamma$ &             Prob. &        $E_\gamma$ &             Prob. \\
     0.062  &      16.4  &     0.511  &       0.0   \\
     0.091  &      53.2  &            &             \\
     0.153  &      30.3  &            &             \\
     0.511  &     186.0  &            &             \\
\hline
\end{tabular}
\end{center}
\end{table}

\begin{table}
\begin{center}
\caption{Cont.
\label{tab_nuc2}}
\vspace{0.3cm}
\begin{tabular}{cccc}
\hline
 \multicolumn{4}{c}{   $\iso{51}Mn \rightarrow \iso{51}Cr \rightarrow \iso{51}V$ } \\
     \multicolumn{2}{c}{$\iso{51}Mn \rightarrow \iso{51}Cr$}  &      \multicolumn{2}{c}{$\iso{51}Cr \rightarrow \iso{51}V$}  \\
\hline
  \multicolumn{2}{c}{$t_{1/2}=$          0.032      d} &  \multicolumn{2}{c}{$t_{1/2}=$         27.700     d} \\
        \multicolumn{2}{c}{$Q_{\gamma}=$          0.992    MeV} &        \multicolumn{2}{c}{$Q_{\gamma}=$          0.032   MeV} \\
        \multicolumn{2}{c}{$Q_{\rm th}=$          0.933    MeV} &        \multicolumn{2}{c}{$Q_{\rm th}=$          0.000   MeV} \\
\hline
       $E_\gamma$ &             Prob. &        $E_\gamma$ &             Prob. \\
     0.511  &     194.2  &     0.320  &       9.9   \\
\hline
 \multicolumn{4}{c}{  $\iso{55}Co \rightarrow \iso{55}Fe \rightarrow \iso{55}Mn$ } \\
     \multicolumn{2}{c}{$\iso{55}Co \rightarrow \iso{55}Fe$}  &     \multicolumn{2}{c}{$\iso{55}Fe \rightarrow \iso{55}Mn$}  \\
\hline
  \multicolumn{2}{c}{$t_{1/2}=$          0.730      d} &  \multicolumn{2}{c}{$t_{1/2}=$       1002.200     d} \\
        \multicolumn{2}{c}{$Q_{\gamma}=$          1.943    MeV} &        \multicolumn{2}{c}{$Q_{\gamma}=$          0.000   MeV} \\
        \multicolumn{2}{c}{$Q_{\rm th}=$          0.430    MeV} &        \multicolumn{2}{c}{$Q_{\rm th}=$          0.000   MeV} \\
\hline
       $E_\gamma$ &             Prob. &        $E_\gamma$ &             Prob. \\
     0.477  &      20.2  &     0.511  &       0.0   \\
     0.511  &     152.0  &            &             \\
     0.931  &      75.0  &            &             \\
     1.317  &       7.1  &            &             \\
     1.370  &       2.9  &            &             \\
     1.408  &      16.9  &            &             \\
\hline
 \multicolumn{4}{c}{   $\iso{37}K \rightarrow \iso{37}Ar \rightarrow \iso{37}Cl$ } \\
      \multicolumn{2}{c}{$\iso{37}K \rightarrow \iso{37}Ar$}  &     \multicolumn{2}{c}{$\iso{37}Ar \rightarrow \iso{37}Cl$}  \\
\hline
  \multicolumn{2}{c}{$t_{1/2}=$          0.000      d} &  \multicolumn{2}{c}{$t_{1/2}=$         35.040     d} \\
        \multicolumn{2}{c}{$Q_{\gamma}=$          1.072    MeV} &        \multicolumn{2}{c}{$Q_{\gamma}=$          0.000   MeV} \\
        \multicolumn{2}{c}{$Q_{\rm th}=$          2.347    MeV} &        \multicolumn{2}{c}{$Q_{\rm th}=$          0.000   MeV} \\
\hline
       $E_\gamma$ &             Prob. &        $E_\gamma$ &             Prob. \\
     0.511  &     199.8  &     0.003  &       5.5   \\
     2.796  &       1.8  &            &             \\
     3.601  &       0.0  &            &             \\
\hline
 \multicolumn{4}{c}{  $\iso{52}Fe \rightarrow \iso{52}Mn \rightarrow \iso{52}Cr$ } \\
     \multicolumn{2}{c}{$\iso{52}Fe \rightarrow \iso{52}Mn$}  &     \multicolumn{2}{c}{$\iso{52}Mn \rightarrow \iso{52}Cr$}  \\
\hline
  \multicolumn{2}{c}{$t_{1/2}=$          0.345      d} &  \multicolumn{2}{c}{$t_{1/2}=$          0.015     d} \\
        \multicolumn{2}{c}{$Q_{\gamma}=$          0.751    MeV} &        \multicolumn{2}{c}{$Q_{\gamma}=$          2.447   MeV} \\
        \multicolumn{2}{c}{$Q_{\rm th}=$          0.191    MeV} &        \multicolumn{2}{c}{$Q_{\rm th}=$          1.113   MeV} \\
\hline
       $E_\gamma$ &             Prob. &        $E_\gamma$ &             Prob. \\
     0.169  &      99.2  &     0.511  &     190.0   \\
     0.378  &       1.6  &     1.434  &      98.3   \\
     0.511  &     112.0  &     2.965  &       1.0   \\
            &            &     3.129  &       1.0   \\
\hline
 \multicolumn{4}{c}{  $\iso{44}Ti \rightarrow \iso{44}Sc \rightarrow \iso{44}Ca$ } \\
     \multicolumn{2}{c}{$\iso{44}Ti \rightarrow \iso{44}Sc$}  &     \multicolumn{2}{c}{$\iso{44}Sc \rightarrow \iso{44}Ca$}  \\
\hline
  \multicolumn{2}{c}{$t_{1/2}=$      21915.000      d} &  \multicolumn{2}{c}{$t_{1/2}=$          0.165     d} \\
        \multicolumn{2}{c}{$Q_{\gamma}=$          0.000    MeV} &        \multicolumn{2}{c}{$Q_{\gamma}=$          2.136   MeV} \\
        \multicolumn{2}{c}{$Q_{\rm th}=$          0.000    MeV} &        \multicolumn{2}{c}{$Q_{\rm th}=$          0.596   MeV} \\
\hline
       $E_\gamma$ &             Prob. &        $E_\gamma$ &             Prob. \\
     0.511  &       0.0  &     0.511  &     188.5   \\
            &            &     1.157  &      99.9   \\
\hline
\end{tabular}
\end{center}
\end{table}

\begin{table}
\begin{center}
\caption{Same as Table~\ref{tab_nuc1}, but now for 1-step decays.
\label{tab_nuc3}}
\vspace{0.3cm}
\begin{tabular}{cccc}
\hline
    \multicolumn{4}{c}{$\iso{41}Ar \rightarrow \iso{41}K$  }   \\
\hline
  \multicolumn{4}{c}{$t_{1/2}=$          0.076     d} \\
        \multicolumn{4}{c}{$Q_{\gamma}=$          1.283   MeV} \\
        \multicolumn{4}{c}{$Q_{\rm th}=$          0.464   MeV} \\
\hline
     & $E_\gamma$ &          Prob. &  \\
 &      1.294  &      99.2     & \\
 &      1.677  &       0.1     & \\
\hline
    \multicolumn{4}{c}{$\iso{42}K \rightarrow \iso{42}Ca$  }   \\
\hline
  \multicolumn{4}{c}{$t_{1/2}=$          0.513     d} \\
        \multicolumn{4}{c}{$Q_{\gamma}=$          0.276   MeV} \\
        \multicolumn{4}{c}{$Q_{\rm th}=$          1.430   MeV} \\
\hline
     & $E_\gamma$ &          Prob. &  \\
 &      1.525  &      18.1     & \\
\hline
    \multicolumn{4}{c}{$\iso{43}K \rightarrow \iso{43}Ca$  }   \\
\hline
  \multicolumn{4}{c}{$t_{1/2}=$          0.929     d} \\
        \multicolumn{4}{c}{$Q_{\gamma}=$          0.964   MeV} \\
        \multicolumn{4}{c}{$Q_{\rm th}=$          0.310   MeV} \\
\hline
     & $E_\gamma$ &          Prob. &  \\
 &      0.221  &       4.8     & \\
 &      0.373  &      86.8     & \\
 &      0.397  &      11.9     & \\
 &      0.404  &       0.4     & \\
 &      0.593  &      11.3     & \\
 &      0.618  &      79.2     & \\
 &      0.801  &       0.1     & \\
 &      0.990  &       0.3     & \\
 &      1.022  &       2.0     & \\
 &      1.394  &       0.1     & \\
\hline
   \multicolumn{4}{c}{$\iso{43}Sc \rightarrow \iso{43}Ca$  }   \\
\hline
  \multicolumn{4}{c}{$t_{1/2}=$          0.162     d} \\
        \multicolumn{4}{c}{$Q_{\gamma}=$          0.985   MeV} \\
        \multicolumn{4}{c}{$Q_{\rm th}=$          0.419   MeV} \\
\hline
     & $E_\gamma$ &          Prob. &  \\
 &      0.373  &      22.5     & \\
 &      0.511  &     176.2     & \\
\hline
   \multicolumn{4}{c}{$\iso{47}Sc \rightarrow \iso{47}Ti$  }   \\
\hline
  \multicolumn{4}{c}{$t_{1/2}=$          3.349     d} \\
        \multicolumn{4}{c}{$Q_{\gamma}=$          0.109   MeV} \\
        \multicolumn{4}{c}{$Q_{\rm th}=$          0.162   MeV} \\
\hline
     & $E_\gamma$ &          Prob. &  \\
 &      0.159  &      68.3     & \\

\hline
   \multicolumn{4}{c}{$\iso{61}Co \rightarrow \iso{61}Ni$  }   \\
\hline
  \multicolumn{4}{c}{$t_{1/2}=$          0.069     d} \\
        \multicolumn{4}{c}{$Q_{\rm th}=$          0.459   MeV} \\
\hline
     & $E_\gamma$ &          Prob. &  \\
 &      0.067  &      84.7     & \\
 &      0.842  &       0.8     & \\
 &      0.909  &       3.6     & \\
\hline
\end{tabular}
\end{center}
\end{table}

\label{lastpage}

\end{document}